\documentclass[twocolumn,english,aps,prl,superscriptaddress]{revtex4-1}
\usepackage[T1]{fontenc}
\usepackage[latin9]{inputenc}
\usepackage{color}
\usepackage{amsmath}
\usepackage{amssymb}
\usepackage{graphicx}
\usepackage{natbib}
\usepackage{braket}
\usepackage{psfrag}
\usepackage{tikz}
\usepackage{physics}
\usepackage{cancel}
\usepackage{nccmath}
\usepackage{qcircuit}
\usepackage{enumitem}

\usepackage[colorlinks=true,urlcolor=blue,citecolor=blue,linkcolor=blue]{hyperref}

\usepackage{babel}

\usepackage[normalem]{ulem}

\begin{document}

\title{Benchmarking the noise sensitivity of different parametric two-qubit gates in a single superconducting quantum computing platform}

\author{M. Ganzhorn}
\affiliation{IBM Research Zurich, S\"aumerstrasse 4, 8803 R\"uschlikon, Switzerland}
\author{G. Salis}
\affiliation{IBM Research Zurich, S\"aumerstrasse 4, 8803 R\"uschlikon, Switzerland}
\author{D.~J. Egger}
\affiliation{IBM Research Zurich, S\"aumerstrasse 4, 8803 R\"uschlikon, Switzerland}
\author{A. Fuhrer}
\affiliation{IBM Research Zurich, S\"aumerstrasse 4, 8803 R\"uschlikon, Switzerland}
\author{M. Mergenthaler}
\affiliation{IBM Research Zurich, S\"aumerstrasse 4, 8803 R\"uschlikon, Switzerland}
\author{C. M\"uller}
\affiliation{IBM Research Zurich, S\"aumerstrasse 4, 8803 R\"uschlikon, Switzerland}
\author{P. M\"uller}
\affiliation{IBM Research Zurich, S\"aumerstrasse 4, 8803 R\"uschlikon, Switzerland}
\author{S. Paredes}
\affiliation{IBM Research Zurich, S\"aumerstrasse 4, 8803 R\"uschlikon, Switzerland}
\author{M. Pechal}
\affiliation{IBM Research Zurich, S\"aumerstrasse 4, 8803 R\"uschlikon, Switzerland}
\author{M. Werninghaus}
\affiliation{IBM Research Zurich, S\"aumerstrasse 4, 8803 R\"uschlikon, Switzerland}
\author{S. Filipp}
\affiliation{IBM Research Zurich, S\"aumerstrasse 4, 8803 R\"uschlikon, Switzerland}

\begin{abstract}
The possibility to utilize different types of two-qubit gates on a single quantum computing platform adds flexibility in the decomposition of quantum algorithms. A larger hardware-native gate set may decrease the number of required gates, provided that all gates are realized with high fidelity. Here, we benchmark both controlled-Z (CZ) and exchange-type (iSWAP) gates using a parametrically driven tunable coupler that mediates the interaction between two superconducting qubits.  Using randomized benchmarking protocols we estimate an error per gate of  $0.9\pm0.03\%$ and $1.3\pm0.4\%$ fidelity for the CZ and the iSWAP gate, respectively. We argue that spurious $ZZ$-type couplings are the dominant error source for the iSWAP gate, and that phase stability of all microwave drives is of utmost importance. Such differences in the achievable fidelities for different two-qubit gates have to be taken into account when mapping quantum algorithms to real hardware. 
\end{abstract}

\date{\today}

\maketitle

\section{Introduction}

With noisy quantum computers \cite{Preskill2018} it is important to complete a quantum calculation or simulation within the available coherence time. To reach this goal, high-fidelity gate operations, high qubit-qubit connectivity and the ability to carry out operations on multiple qubit patches in parallel are essential \cite{Cross2019}. Moreover, better results are obtained when the device architecture is tailored to the quantum algorithm \cite{Schuch2003,Abrams2019}. For example, in chemistry calculations the number of particle excitations needs to be preserved which makes the iSWAP gate the optimal choice \cite{Barkoutsos2018,Ganzhorn2019}. For quantum approximate optimization algorithms, on the other hand, a controlled phase gate is better matched  to the computational task \cite{Otterbach2017,Bengtsson2019}. Ideally, the quantum computing hardware supports a gate set with multiple types of single-qubit and two-qubit operations. Different types of single-qubit operations can often be realized by choosing a suitable amplitude, phase and time of a control pulse. In contrast, the nature of two-qubit operations depends on the available interactions and the control capabilities of the architecture. 

For superconducting qubits there are several options to realize two-qubit gates. One is the use of microwave drives applied to fixed-frequency qubits. Examples are the cross-resonance gate \cite{Rigetti2010,Chow2011,Sheldon2016b} based on a controlled-rotation (ZX-)type interaction, the bSWAP gate \cite{Poletto2012} based on an bi-photon $\rm{XX}\!-\!\rm{YY}$-type interaction and gates that directly involve microwave cavity states \cite{Leek2009,Paik2016,Egger2018b}. 
With frequency-tunable qubits, both iSWAP gates based on resonant exchange ($\rm{XX}\!+\!\rm{YY}$) type interactions \cite{Dewes2012,Neill2017} as well as controlled-Z (CZ) gates based on ZZ-type interactions \cite{Strauch2003,DiCarlo2009} can be realized by tuning the qubits or their higher excited states close to resonance.
Such gates can elegantly be realized also parametrically by modulating the frequency of a single qubit  \cite{Beaudoin2012,Strand2013,Didier2018,Caldwell2018,Abrams2019}.
 
To avoid negative effects of additional flux-noise on frequency-tunable qubits, fixed-frequency qubits can be combined with tunable couplers (TC). A broad range of two-qubit gates can then be engineered via a parametric modulation of the coupling \cite{Bertet2006,Ploeg2007,Chen2014a,McKay2016,Roth2017,Roushan2017}. Different types of interactions, such as iSWAP exchange-type ($\rm{XX}\!+\!\rm{YY})$, bSWAP ($\rm{XX}\!-\!\rm{YY}$) or controlled-phase ZZ-type gates are possible simply by choosing the correct modulation frequency \cite{Harrabi2009,Roth2017,Ganzhorn2019,Krantz2019}. A multi-tone modulation even allows for combinations of these interactions simultaneously \cite{Roth2019}, and with an analog control of the modulation amplitude, adiabatic protocols can be implemented \cite{Salis2020}.

To benefit from an extended gate set, all gates must be executed with high fidelity. While the tunable-coupler supports both fast iSWAP and CZ gates on the same platform with almost identical hardware requirements, gate-fidelities around $99\%$ have been reached with the CZ gate \cite{Bengtsson2019}, but typical iSWAP fidelities remain lower \cite{McKay2016,Ganzhorn2019,Han2020}. Here, we explore both types of gates on the same device,  characterize their respective fidelities and analyze the specific sensitivity of the gates to the various sources of coherent and incoherent errors.  

\begin{figure}[!htbp]
\centering
\includegraphics[width = 86mm]{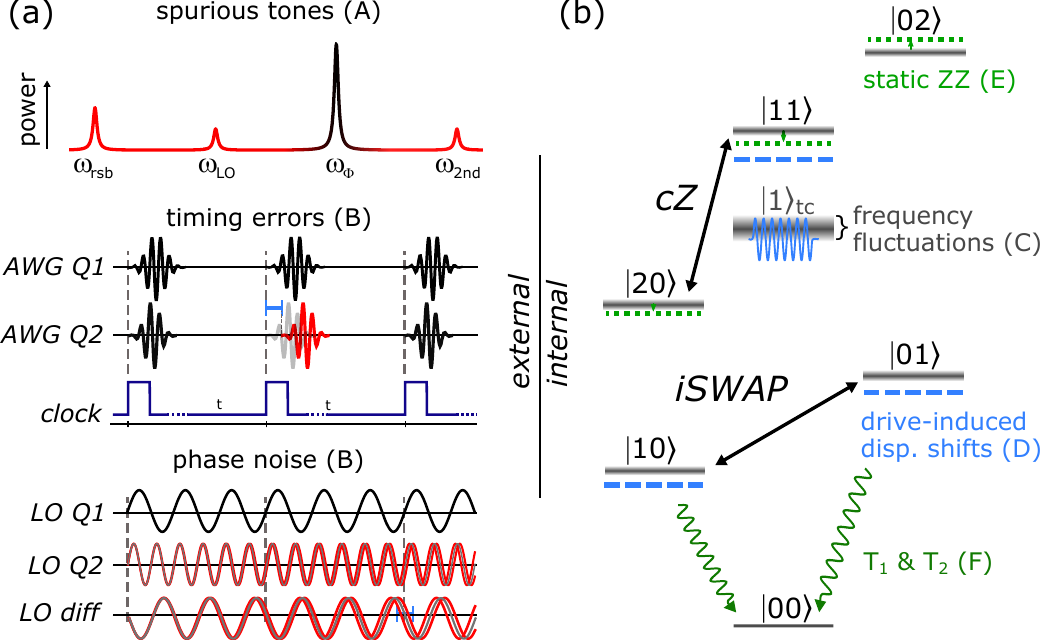}
\caption{\label{fig:setup} 
Error sources split into (a) external control errors and (b) internal system-related errors. (Type A) The up-conversion mixing process leads to spurious signals at integer multiples of the sideband frequency ($\omega_{\rm LO}$, $\omega_{\rm 2nd}$, $\omega_{\rm rsb}$) in addition to the drive signal at frequency $\omega_{\Phi}$. (Type B) Timing errors in the pulses generated by the arbitrary waveform generators (AWGs) and phase noise of the local oscillator (LO) sources lead to phase errors, e.~g. in the frame rotating at the qubit difference frequency. (b)
Level diagram of the system illustrating internal, device-level errors. $\ket{n_1 n_2}$ ($n_{1,2} \in \left\{0,1,2\right\}$) denotes the two-qubit state  with the TC in its ground state.
Errors are caused by fluctuations of the qubit frequencies due to frequency fluctuations of the coupler (type C),  by drive-induced dispersive shifts of the qubit frequencies (type D) and ZZ-type interactions (type E) related to shifts of the $\ket{11}$ state, e.~g. due to the presence of higher excited qubit states $\ket{20}$ and $\ket{02}$.  Other intrinsic decoherence mechanism such as dissipation and decoherence (type F) are illustrated for the lowest qubit levels.}
\end{figure}

We specifically consider the following types of errors as illustrated in Figure \ref{fig:setup}:
\begin{enumerate}[label=(\Alph*),start=1]
\itemsep-4pt
    \item errors caused by spurious coherent tones that stem from uncontrolled harmonics in the signal generation,
    \item relative phase errors caused by random pulse delays and phase noise in the drive signals,
    \item relative phase errors caused by coupler-induced frequency drifts of the qubits,
    \item drive-induced dispersive shifts during the gate caused by the coupling between qubits and the modulated TC, 
    \item static $ZZ$-type errors caused by the interaction between $\ket{11}$ and $\ket{20}$ states \cite{Mundada2019,Han2020}, and
    \item intrinsic dissipation and decoherence ($T_1$ and $T_2$).
\end{enumerate}
Type A and B are external errors that occur during pulse generation [Fig.~\ref{fig:setup}(a)]. Type C, D and E are internal errors related to the quantum system itself [see Fig.~\ref{fig:setup}(b)], and add to intrinsic energy relaxation and decoherence processes (type F). 

As we will show in the following, the iSWAP gate is in particular susceptible to phase errors and ZZ-type crosstalk, in contrast to the CZ gate which is resilient to phase errors and for which ZZ-type errors can be avoided by proper calibration.

%

\section{Description of the setup}

We use two fixed-frequency transmon qubits $Q_1$ and $Q_2$ with frequencies $\omega_{10}/2\pi=5.089~\rm{GHz}$ and $\omega_{01}/2\pi = 6.189~\rm{GHz}$. Both of them are capacitively coupled to a common TC  (see \emph{Supplementary Material}). The device layout is similar to the one used in Ref.~\cite{Ganzhorn2019}. 
The TC frequency 
\begin{equation}
\label{eq:TCfreq}
\omega_{\textrm c}(t) = \omega_{\textrm c}^{0} \sqrt{\gamma(t)\abs{\cos(\pi\Phi(t)/\Phi_0)}}   
\end{equation}  
with maximum frequency $\omega_{\textrm c}^{0}/2\pi = 8.1~\rm{GHz}$ is modulated by applying an oscillating magnetic flux 
\begin{equation}
\label{eq:Fluxmodulation}
\Phi(t) = \Phi_{\text{DC}} + \delta_{\Phi} \cos(\omega_\Phi t + \eta) 
\end{equation} 
with variable phase $\eta$ to the SQUID-loop of the TC. $\gamma(t)=\{1+d^2\tan^2(\pi\Phi(t)/\Phi_0)\}^{1/2}$ takes the asymmetry $d$ of the SQUID-loop into account \cite{Koch2007}. Depending on the drive frequency $\omega_\Phi$, the frequency modulation induces transitions between different energy levels \cite{McKay2016, Roth2017}. We consider the CZ and the iSWAP gate.
The iSWAP gate is activated by setting $\omega_\Phi$ to the qubits' difference frequency $\omega_\Delta = \omega_{01}-\omega_{10}$~\cite{McKay2016}.
A CZ gate is implemented by choosing $\omega_{\Phi}=\omega_\alpha \equiv \omega_{11}-\omega_{20}=\omega_\Delta-\alpha_1$ [Fig.~\ref{fig:setup}(b)] which drives the transition between the  $\ket{20}$ and $\ket{11}$ state \cite{Strauch2003,Filipp2019,Krantz2019}. $\omega_{20}$ denotes the frequency of the second excited state of qubit $Q_1$ and $\alpha_1/2\pi= -310~\rm{MHz}$ its anharmonicity. An in-phase/quadrature (IQ) mixer is used to generate the microwave control pulses: A low-frequency pulse from an arbitrary waveform generator (AWG) is single-sideband modulated onto a carrier signal generated by a vector signal sources to create pulses with adjustable frequency, amplitude and phase. 

\section{Characterization of the CZ gate}

\begin{figure}[!htbp]
\centering
\includegraphics[width =86mm]{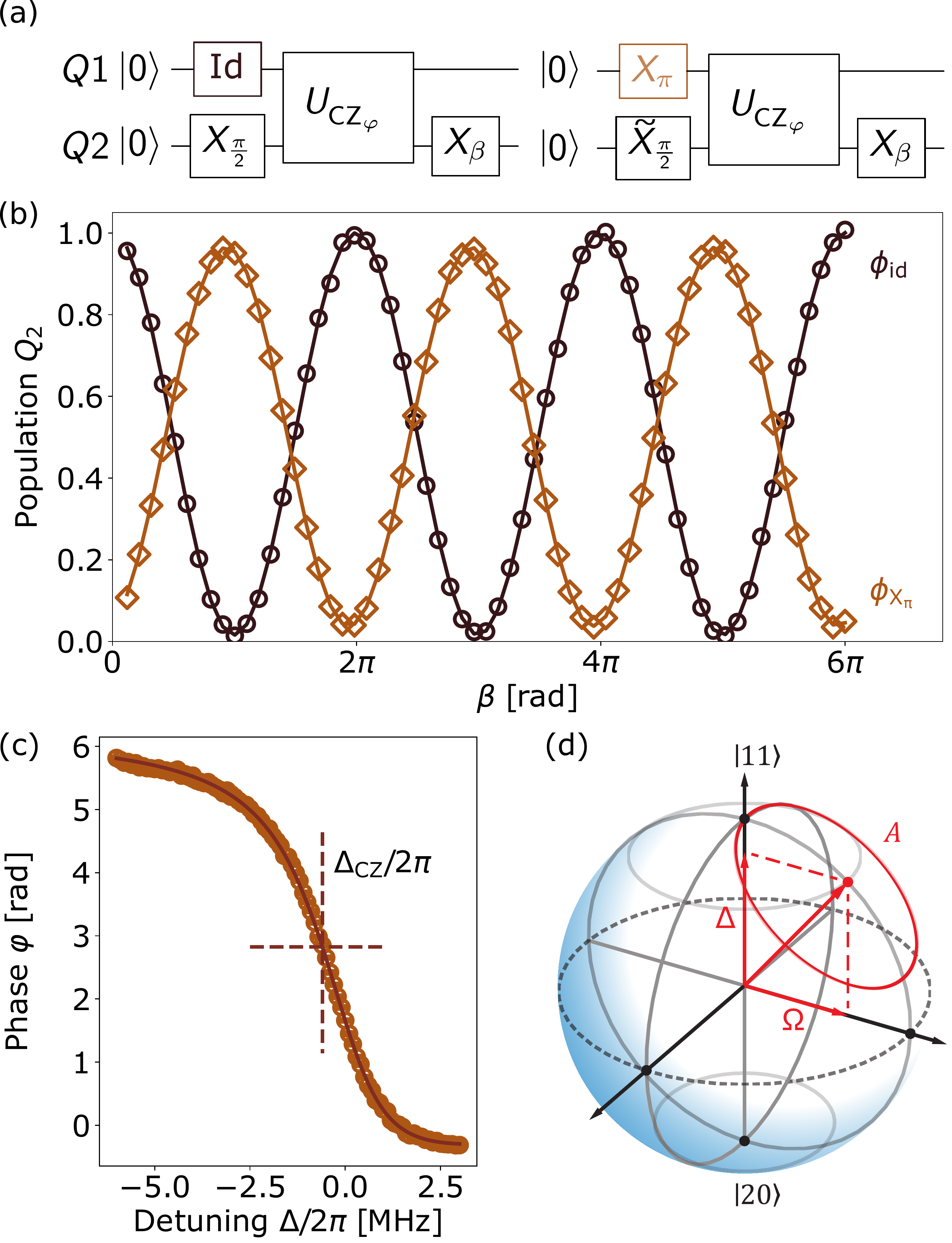}
\caption{Calibration of the CZ gate. (a) Schematic of a Ramsey-type experiment measuring the phase on qubit $Q_2$ with an interleaved CZ$_{\varphi}$ gate $U_{{\rm CZ}_\varphi}$ with qubit $Q1$ in its ground state (left) or its excited state (right). (b) Population in qubit $Q_2$ measured for different azimuthal rotation angles $\beta$ on the trailing $\tilde{X}_{\pi/2}$ pulses in (a). Grey (brown) Ramsey fringes are obtained with the left (right) quantum circuit in (a) and a sinusoidal fit yields the phase $\phi_{\rm Id}$ ($\phi_{X_{\pi}}$). (c) Phase $\varphi=\phi_{\pi}-\phi_{X_{\rm Id}}$ as a function of the detuning $\Delta$ for a sideband frequency of $\omega_{sb}/2\pi = 105$ MHz and $\tau_{\rm gate}=591~\rm{ns}$. The solid line is a fit based on Eq.~(\ref{eq:geom}), the dashed cross indicates the detuning $\Delta_{\rm CZ}$ at which the best RB gate fidelity is measured. (d) Bloch sphere spanned by $\ket{11}$ and $\ket{20}$. The phase $\phi_g$ is determined by the solid angle $|A| = 2\phi_g$ enclosed by the path of the state vector. \label{fig:cphase}}
\end{figure}

By modulating the TC at frequency $\omega_\alpha$, a resonance condition between the states $\ket{20}$ and $\ket{11}$ is established in a reference frame rotating at the respective qubit frequencies, similar to bringing the energies of these levels into or close to resonance by directly tuning the qubit frequencies \cite{Strauch2003,DiCarlo2009,Egger2013a,Martinis2014,Rol2019}. When starting in the $\ket{11}$ state, this leads to oscillations of the population between the $\ket{11}$ and the $\ket{20}$ state. At a given TC modulation amplitude, the length of the modulation pulse (gate length $\tau_{\rm gate}$) is chosen such that one full oscillation occurs, i.e. such that the state population returns to $\ket{11}$. The state then acquires a phase $\varphi$, thus implementing a CZ gate. The dynamics is described by the unitary
\begin{align}
\label{eq:UCZphi}
U_{\rm CZ}(\varphi) & = 
\begin{pmatrix} 
1 & 0 & 0 & 0 \\
0 & 1 & 0 & 0 \\ 
0 & 0 & 1 & 0 \\ 
0 & 0 & 0 & e^{-i\varphi} \\ 
\end{pmatrix},
\end{align}
provided that leakage into the $\ket{20}$ state is avoided by a suitable pulse shape.  The acquired phase, $\varphi = \phi_g + \phi_\zeta$, is composed of a geometric component  $\phi_g$ and an energy-dependent dynamical component $\phi_\zeta$. The geometric phase $\phi_g$ is proportional to the solid angle enclosed by the evolution of the state vector on the two-dimensional Bloch sphere \cite{Sjoqvist2015} spanned by $\ket{11}$ and $\ket{20}$, see Fig.\ref{fig:cphase}\textcolor{red}{(d)}. It depends on the ratio between the effective coupling strength $\Omega$ of the $\ket{11}\leftrightarrow\ket{20}$ transition and the detuning $\Delta = \omega_\Phi - \omega_\alpha$ as
\begin{equation}
\varphi_g = \pi \left\{1 - \cos \left[ \arctan \left(\frac{\Omega}{\Delta}\right)\right]\right\},
\label{eq:geom}
\end{equation}
where $\tau = 2\pi/\sqrt{\Omega^2 + (\Delta-\zeta)^2}$ is the duration of the gate~\cite{Leek2009}.
The dynamical phase component $\phi_\zeta$ is induced by the frequency shift of the $\ket{11}$ state by its static coupling to nearby $\ket{20}$ and $\ket{02}$ states and is described by $\phi_\zeta = \zeta \cdot \tau$, with the pulse duration  $\tau$  and the  ZZ-type shift $\zeta= \omega_{11} - \omega_{01} - \omega_{10}$ of the $\ket{11}$ state given by the Hamiltonian 

\begin{eqnarray}
H_{\zeta}/\hbar = \zeta \ket{11}\bra{11} = \frac{\zeta}{4} \left({\rm ZZ}\!-\!\rm{IZ}\!-\!\rm{ZI}\!+\!\rm{II} \right)
\label{eq:heff_cphase}
\end{eqnarray}

with I denoting the single-qubit identity operation and Z the $\sigma_z$ Pauli-operator. 
To measure the total controlled phase $\varphi$ we compare the phases of the superposition states $\ket{00}+\ket{01}$ and $\ket{10}+\ket{11}$ after application of the CZ gate. For this, a Ramsey-type experiment on qubit $Q_2$ with an interleaved CZ gate is performed with qubit $Q_1$ being initialized either in its ground state or in its excited state [Fig. \ref{fig:cphase}(a)]. The Ramsey fringes are measured by varying the angle $\beta$ of virtual $Z$ rotations~\cite{McKay2017} before the final $X_{\pi/2}$ pulse (implemented experimentally by varying the phase of the $X_{\pi/2}$ pulse), see Fig.~\ref{fig:cphase}(b). The phases $\phi_{\rm Id}$ and $\phi_{X_{\pi}}$ of the measured oscillations in $\beta$ then determine $\varphi =\phi_{X_{\pi}}-\phi_{\rm Id}$. The procedure is repeated for different detuning $\Delta$. The resulting phase $\varphi$ is shown in Figure \ref{fig:cphase}(c).

From a fit of $\phi_g + \phi_\zeta$ [using Eq.~(\ref{eq:geom})] to the measured data we find the value of the ZZ-shift component $\zeta/2\pi = -355\pm 1~\rm{kHz}$ and the drive strength $\Omega/2\pi = 1.450 \pm 0.006~\rm{MHz}$. A detuning $\Delta/2\pi = -770~\rm{kHz}$ results in a total phase shift $\varphi = \pi$, the expected value for an ideal CZ gate. By minimizing the error per gate using a randomized benchmarking (RB) protocol for varying $\Delta$ we however find a slightly different value $\Delta_{\rm CZ}/2\pi = -600~\rm{kHz}$ (where $\varphi=2.82\pm0.09~\rm{rad}$), as discussed in the \emph{Supplementary Material} together with the complete calibration procedure. These values were obtained for $\tau_{\rm gate}=591~\rm{ns}$. 

\begin{figure}[!th]
\centering
\includegraphics[width =86mm]{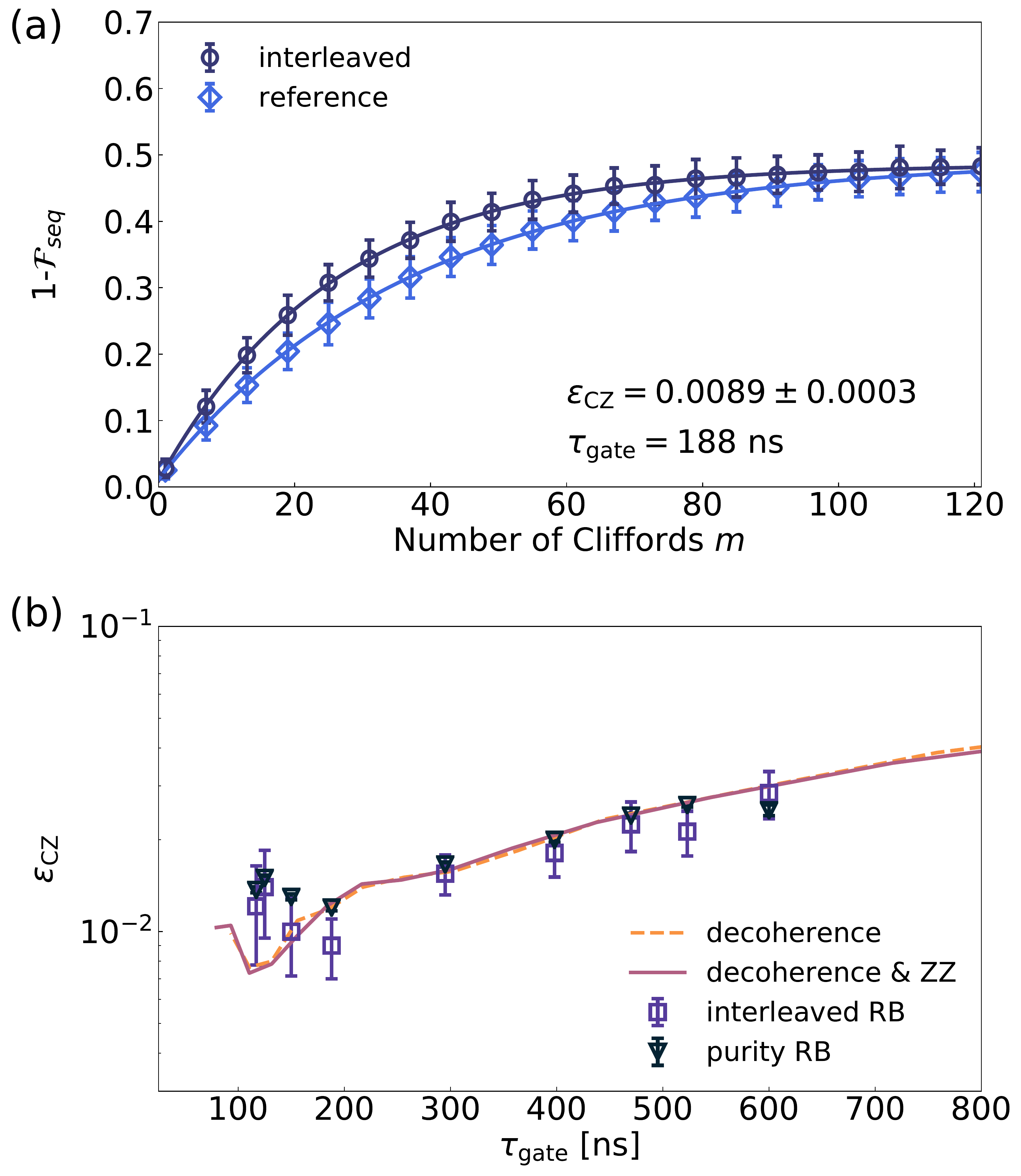}
\caption{Benchmarking of a $CZ$ gate at $\Phi_{\rm DC} = 0.15 \Phi_0$. (a) Interleaved RB at a gate length of $\tau_{\rm gate} = 188~\rm{ns}$. The measurement is performed on qubit $Q_1$, averaged over $200$ realizations of randomized Clifford gate sequences. Similar results are obtained for $Q_2$ (not shown). The blue diamonds and dark blue circles depict the reference and interleaved sequence, respectively. (b) Average error per gate $\epsilon_{\rm CZ}$ (violet squares) and purity error per gate (dark blue triangles) measured as function of the gate length $\tau_{\rm gate}$. Dashed orange lines represent the simulated error per CZ gate including qubit decoherence rates, while solid violet lines depict the simulated error per CZ gate including decoherence and an additional $ZZ$-type crosstalk contribution with $\zeta=-200$\,kHz.\label{fig:epg_cphase}}
\end{figure}

For different TC modulation amplitudes and therefore different $\tau_{\rm gate}$ we repeat the calibration procedure and perform interleaved randomized benchmarking (RB)~\cite{Magesan2012} at $\Delta=\Delta_{\rm CZ}$ to assess the two-qubit gate errors. From the sequence fidelity $\mathcal{F}_{\rm seq}$ of the reference and interleaved RB sequences measured as a function of the length of the Clifford sequence, we determine the average error per gate $\epsilon_{\rm CZ}$ for  $U_\text{CZ}$, see Fig.~\ref{fig:epg_cphase}(a). The average error per gate  $\epsilon_{\rm CZ}$ is shown in Fig.\ref{fig:epg_cphase}(b) as a function of $\tau_{\rm gate}$. We find the lowest error at $\tau_{\rm gate} = 188~\rm{ns}$. The gate error there, $\epsilon_{\rm CZ} = 0.0089 \pm 0.0003$, is a trade-off between decoherence errors and leakage errors~\cite{McKay2016,Roth2017}. 

\section{Characterization of the {i}SWAP-gate}

\begin{figure}[!t]
\centering
\includegraphics[width =86mm]{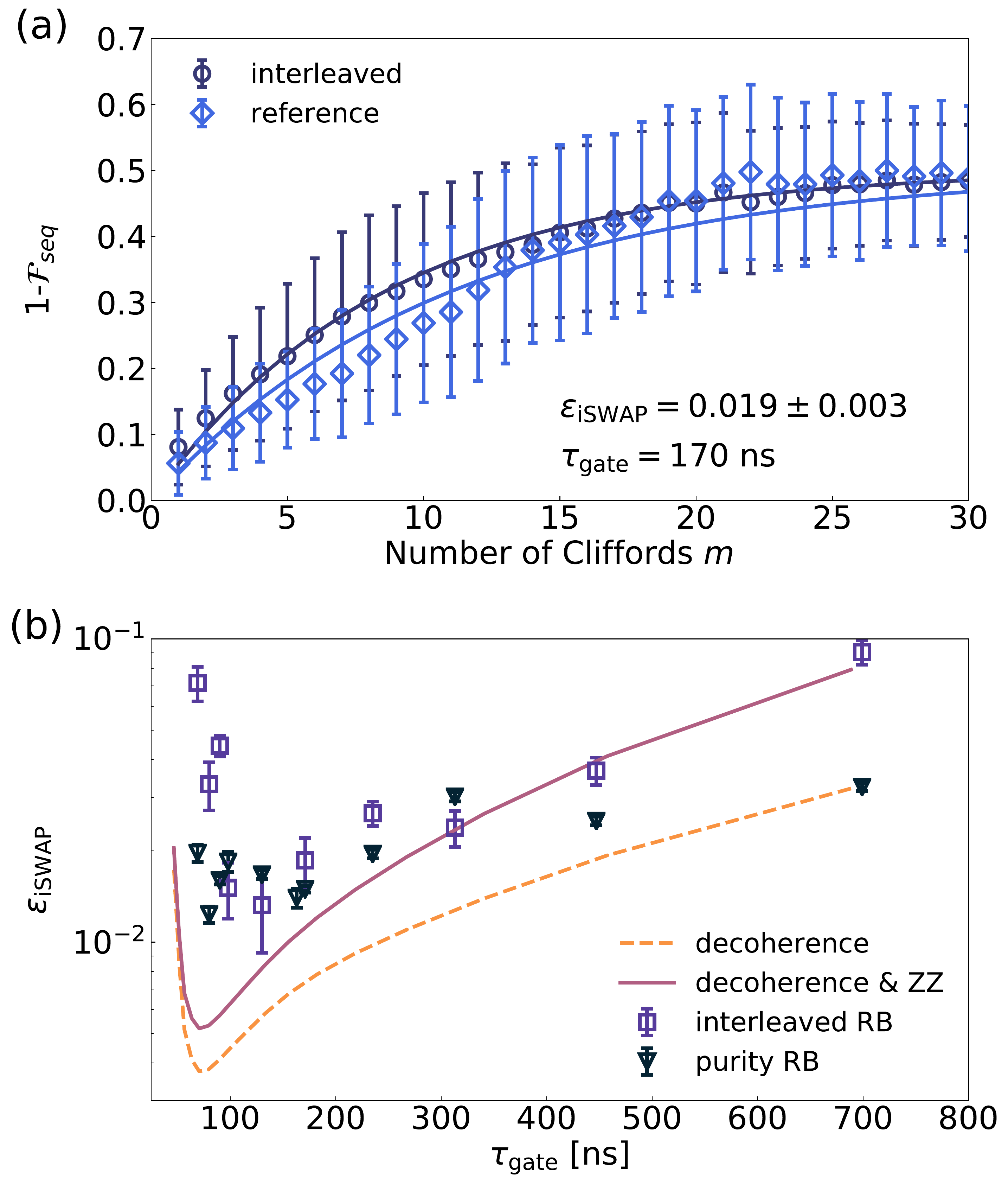}
\caption{Benchmarking of an iSWAP gate at $\Phi_{\rm DC} = 0.15 \Phi_0$. (a) Interleaved RB at a gate length of $\tau_{\rm gate} = 170~\rm{ns}$. The measurement is performed on qubit $Q_1$, averaged over $200$ random RB circuits. Similar results are obtained for $Q_2$ (not shown). The blue diamonds and dark blue circles depict the reference and interleaved sequence, respectively. (b) Average error per gate $\epsilon_{\rm iSWAP}$ (violet squares) and purity error per gate (dark blue triangles) measured as function of the gate length $\tau_{\rm gate}$. The dashed orange line represents the simulated error per iSWAP gate including qubit decoherence, while the solid violet line depicts the simulated error per iSWAP gate including decoherence and an additional ZZ contribution with $\zeta=-200$\,kHz. \label{fig:epg_iswap}}
\end{figure}

The iSWAP gate is realized by modulating the coupler at the qubit difference frequency $\omega_\Delta$ that drives the $\ket{01}\leftrightarrow\ket{10}$ transition at a rate $\Omega$ which is determined by the modulation amplitude \cite{Roth2017}. The effective Hamiltonian is described by
\begin{align}
\label{eq:heff_iswap}
H_{\rm XY}/\hbar=\frac{\Omega}{4} \left[\cos\eta \left(\rm{XX}\!+\!\rm{YY}\right)\!-\!\sin\eta \left(\rm{YX}\!-\!\rm{XY}\right)\right]
\end{align}
 with the set of Pauli operators $\{$I, X, Y, Z$\}$. $\eta$ is the relative phase between the modulation pulse of the TC and the phase difference of the frames rotating at the qubit frequencies. $H_{\rm XY}$ generates the unitary operation 
\begin{align}
\label{eq:Uex}
U_\text{iSWAP}(\theta,\eta) =\begin{pmatrix} 
1 & 0 & 0 & 0 \\
0 & \cos\theta/2 & ie^{i\eta}\sin\frac{\theta}{2} & 0 \\ 
0 & ie^{-i\eta}\sin\theta/2 & \cos\theta/2 & 0 \\ 
0 & 0 & 0 & 1 \\ 
\end{pmatrix},
\end{align}
where the rotation angle $\theta = \Omega \tau$ is controlled by the length $\tau$ of the pulse and set to $\theta=\pi$ to realize an iSWAP gate $U_{\rm iSWAP}=U_{\rm iSWAP}(\pi,0)$. 
To calibrate the gate we follow a similar procedure as for the CZ gate as discussed in the \emph{Supplementary Material}. In contrast to the CZ gate, the calibration of the gate involves an extra step that adjusts the relative phase between the qubits and the TC to $\eta = 0$ via a cross-Ramsey type experiment. This is because the iSWAP-Hamiltonian $H_{\rm XY}$ in Eq.~(\ref{eq:heff_iswap}) depends explicitly on the phase $\eta$ of the drive \cite{Ganzhorn2019}. We then  measure the average error per gate $\epsilon_{\rm iSWAP}$ via the sequence fidelity $\mathcal{F}_{\rm seq}$ of reference and interleaved RB sequences as a function of the number of Cliffords as shown in Fig.~\ref{fig:epg_iswap}(a). At a gate length $\tau_{\rm gate} = 170~\rm{ns}$, similar to the duration of the best CZ gate, we find an error per gate of $0.019\pm0.003$. This value is almost twice as large as the error of the best CZ gate even though we use the same hardware configuration and similar parametric drive frequencies that differ only by the anharmonicity $\alpha_1$. By varying $\tau_{\rm gate}$  we find the minimal error to be $\epsilon_{\rm gate} = 0.0130 \pm 0.004$ at $\tau_{\rm gate} = 130~\rm{ns}$, see Fig.~\ref{fig:epg_iswap}(b). 

\section{Discussion of gate errors}

In addition to the difference in the average error per gate, we notice that the spread of the measured sequence fidelity $\mathcal{F}_{\rm seq}$ for the iSWAP gate is significantly larger as compared to that of the CZ gate. This is seen when comparing the error bars in Fig.~\ref{fig:epg_cphase}(a) with those in Fig. \ref{fig:epg_iswap}(a) and hints to the presence of correlated unitary errors \cite{Ball2016} as will be discussed further below.

To discriminate the different error contributions, we have also measured the purity error determined by the length of the Bloch vector \cite{Wallman2015,Feng2016}. A reduced length of the Bloch vector indicates incoherent errors (e.g. decoherence or thermal population), while coherent errors only change its orientation, but not its length. We measure the density matrix $\rho_m$ of the final state after an RB sequence as a function of the number of Clifford gates $m$ using state tomography. The purity of the final state is then $\rm{Tr}[\rho^2_m]$, which we model by $(3\gamma^{2m}+1)/4$, where $\gamma$ parameterizes a completely depolarizing noise channel $\rho_0 \mapsto \rho(m) = \gamma^m \rho_0 + (1-\gamma^m) I/d$ with the dimension $d=4$ of the Hilbert space. The survival probability $\gamma$ is determined from a fit of the measured purity decay over the number of Cliffords $m$ to $A \gamma^{2m}+B$.
The (purity) error per gate is then given by $\epsilon = (1 - \gamma^{2/3})(d-1)/d = 3(1-\gamma^{2/3})/4$ \cite{Magesan2012a}, where the exponent $2/3$ comes from the average number of $1.5$ iSWAP or CZ operations composing one Clifford gate \cite{Schuch2003}.

We compare the measured errors with  numerical simulations based on a Lindblad-type master equation (in QuTip \cite{Johansson2013a}) taking dissipation and dephasing ($T_1$ and $T_2^*$) into account. The qubits and the TC are modelled with three anharmonic energy levels each, the TC frequency is modulated with an oscillating flux according to Eqs.~(\ref{eq:TCfreq}) and (\ref{eq:Fluxmodulation}).  The average error per gate [Fig.~\ref{fig:epg_cphase}(b) and Fig.~\ref{fig:epg_iswap}(b); solid lines] is extracted from a quantum process tomography (QPT) using $\bar{\epsilon} = 1-(\rm{Tr}[\chi \chi_0])$, where $\chi$ and $\chi_0$ are the simulated and ideal process matrices describing the state evolution (see \emph{Supplementary Material}). 
To separate decoherence errors from ZZ-type errors we also run simulations in which we artificially add  an interaction that compensates the static $ZZ$-term [dashed line in Fig.~\ref{fig:epg_cphase}(b) and Fig.~\ref{fig:epg_iswap}(b)] emerging from the dispersive coupling of the qubits to the TC.   

For the CZ gate the purity error closely follows the measured  error per gate in the interleaved RB sequence [Fig.\ref{fig:epg_cphase}(b)]  indicating that the CZ gate is limited by incoherent errors only.
Moreover, for this gate the difference between the numerical simulation with and without ZZ compensation is negligible, owing to the fact that the static ZZ-type crosstalk contribution can be fully compensated in the gate calibration procedure. In contrast, for the iSWAP gate the purity error is larger than the error expected from decoherence alone [dashed line in Fig.~\ref{fig:epg_iswap}(b)] indicating other noise sources during the gate. Moreover, for both short and long gate times the measured purity error is smaller than the interleaved RB error, which indicates the presence of additional coherent (unitary) errors. 

To get a better understanding of the effect of errors on the CZ and iSWAP gate we consider the error sources listed in the introduction and illustrated in Fig.~\ref{fig:setup}:
\paragraph{Spurious tones -- type A:} Carrier leakage from the mixer and non-linearities in the signal generation lead to spurious signals at multiples of the sideband frequency that drive unwanted transitions and lead to coherent errors and leakage. These effects can be avoided by using large sideband frequencies. For the CZ gate we have chosen $\omega_{\rm sb}/2\pi = 105~\rm{MHz}$ which suppresses these types of errors as shown in the \emph{Supplementary Material}. This strategy does, however, not work for the iSWAP gate that is sensitive to (type B) phase errors (see below): Such phase errors increase linearly with $\omega_{\rm sb}$ in presence of timing errors in the pulse generation. To balance these error sources, we have chosen $\omega_{\rm sb}/2\pi = 5~\rm{MHz}$ for the iSWAP gate. We attribute the difference between the measured  and the significantly smaller simulated errors for gate lengths $\tau_{\rm gate}\lesssim 300~\rm{ns}$ with a minimal error per gate of $0.006$ for $\tau_{\rm gate} = 80~\rm{ns}$ partly to this type of error. 

\paragraph{Relative phase errors -- type B and C:}
Both timing errors in the pulses from the AWGs and phase noise of the signal sources \cite{Ball2016a} lead to fluctuations of the relative phase of the parametric drive relative to the frame rotating at the CZ or iSWAP transition frequency (type B).  Similarly, frequency fluctuations of the TC (type C) induce dispersive shifts of the qubit frequencies and thereby cause random Z-rotations of the qubits and lead to relative phase errors. In particular, we have observed significantly enhanced frequency fluctuation of the qubits when the parametric drive is on. If such frequency fluctuations occur within the time scale of the RB experiment, they can be regarded as a reduction of the effective $T_2^*$ of the qubits from the value of $50~\mu\rm{s}$ and $27~\mu\rm{s}$ when the parametric drive is  off to an amplitude-dependent value between $4~\mu\rm{s}$ and $8~\mu\rm{s}$ (see \emph{Supplementary Material}). Both types of relative phase errors can be described by terms of the form $H_Z/\hbar =  \delta (ZI - IZ)$, a phase advance in the frame rotating at the difference frequency of the qubits. The CZ gate is insensitive to these errors, since it does not depend on the actual phase of the difference frame. This is formally expressed by the vanishing commutator of the effective  CZ-Hamiltonian $H_{\rm ZZ}$ in Eq.~(\ref{eq:heff_cphase}) with the error Hamiltonian $H_Z$, $\left[H_{\rm CZ},H_Z\right]=0$. In contrast, the iSWAP gate explicitly depends on the relative phase of the drive and is, therefore, affected by noise. The non-vanishing commutator $\left[H_{\rm XY}^0, H_Z\right]\propto \rm{XY}\!-\!\rm{YX}$ corresponds to a rotation in the $\ket{01}-\ket{10}$ subspace of the iSWAP Hamiltonian $H_{\rm XY}^0/\hbar = \Omega\,(\rm{XX}\!+\!\rm{YY})/4$ in Eq.~(\ref{eq:heff_iswap}) and, therefore, to errors in the phase $\eta$ of the iSWAP gate. When taking these relative phase errors into account in the numerical simulations,  the larger gate errors for gate lengths $\tau_{\rm gate}\lesssim 300~\rm{ns}$  can be explained, see \emph{Supplementary Material}.

\paragraph{Dispersive shifts -- type D:} For both types of gates, dispersive shifts of the transition frequencies $\ket{01} \leftrightarrow \ket{10}$ during the gate are taken into account by calibrating both the drive frequency and the phase accumulated by the qubits during the pulse as discussed in the \emph{Supplementary Material}. Shifts of the $\ket{11} \leftrightarrow \ket{20}$ transition frequency can, however, only be compensated for the CZ gate as discussed below.
\paragraph{ZZ-type errors -- type E:} The effect of the static ZZ term in Eq.~(\ref{eq:heff_cphase}) can be completely omitted for the CZ gate by adjusting the frequency and length of the gate pulse to compensate the shift of the $\ket{11}$ level. For the iSWAP gate the extra ZZ-term adds undesired controlled-phase type interactions that increase the average gate error, as it becomes evident from numerical simulations: In the ZZ non-compensated simulations we find good agreement with experimental values for longer gate duration $\tau_{\rm gate}>400~\rm{ns}$ [Fig.~\ref{fig:epg_iswap}(b); solid  line] implying that in addition to dissipation and decoherence the error per gate is limited by static ZZ-type crosstalk. Compensating the static ZZ coupling in the numerical simulation leads to an overall decrease of the error per gate to a similar value as the one reached in the purity RB measurement for longer gates [Fig.~\ref{fig:epg_iswap}(b); dashed line]. 
The increase of the error for $\tau_{\rm gate}$ smaller than 100\,ns is attributed to the spectral broadening of the parametric TC pulse that excites unwanted transitions, an effect that could be mitigated using pulse optimization \cite{Werninghaus2020,Motzoi2009}.

\begin{figure}[!htbp]
\centering
\includegraphics[width=86mm]{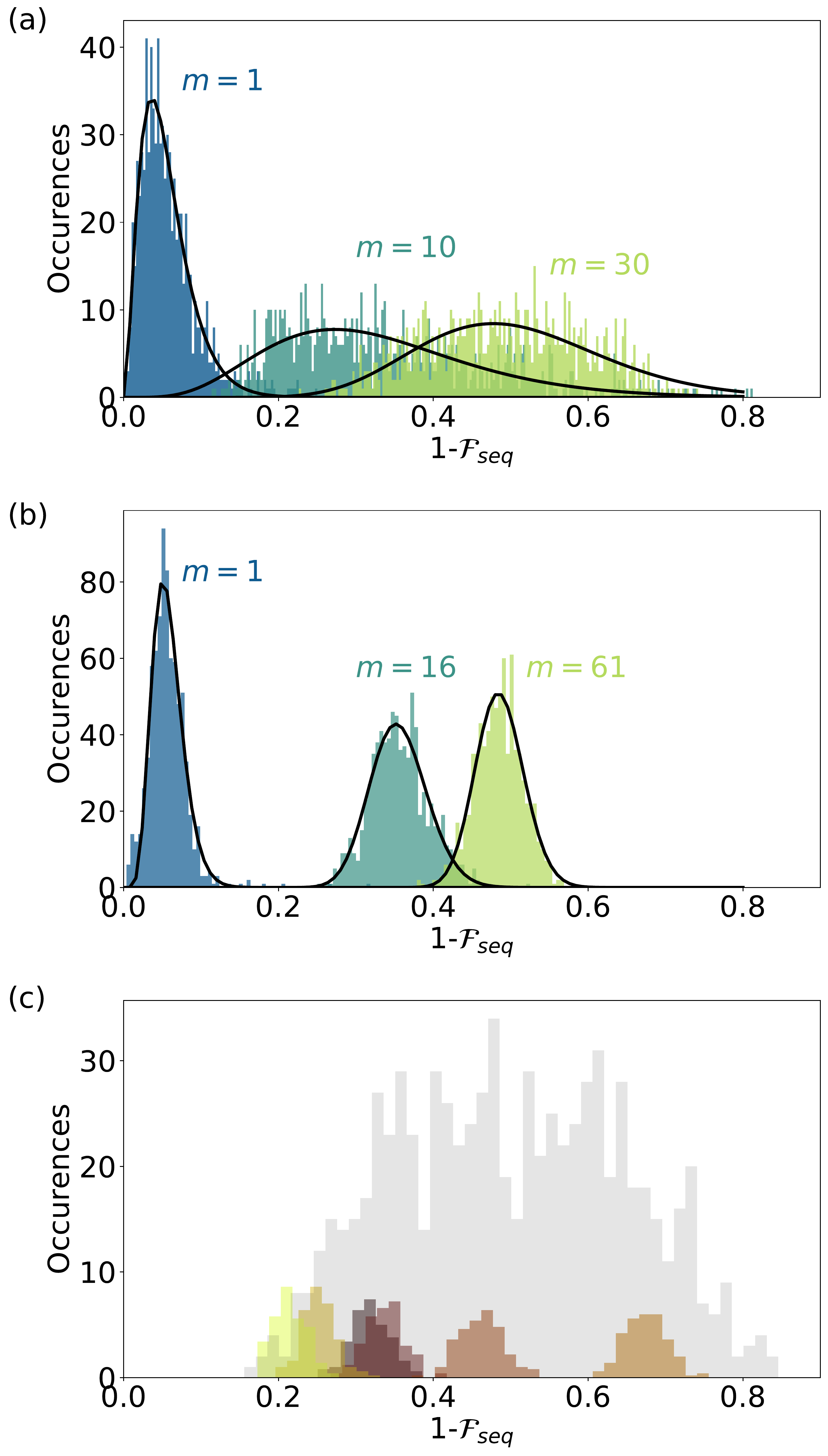}
\caption{Distribution of sequence fidelities for the (a) iSWAP and (b) CZ gate for different number of Cliffords in a standard RB measurement. 800 different realizations have been recorded. Solid lines indicate fits to a $\Gamma$-function \cite{Ball2016}. In (c) the statistical distribution of the sequence fidelity for different realizations with $m=8$ Clifford gates is shown. Each colored histogram shows 150 measurements of a single realization and re-scaled by a factor of 1/5 for better visibility. The grey histogram shows a total of 800 randomization measured one time each. \label{fig:RBdist}}
\end{figure}

\paragraph{Effect of ZZ-type errors.} To further analyze the effect of ZZ terms we analyze histograms of measured $\mathcal{F}_{\rm seq}$ for the CZ and the iSWAP gates for 800 RB realizations. We observe a significant increase of the spread in $\mathcal{F}_{\rm seq}$ with the number of Clifford gates $m$ for the iSWAP gate shown in Fig.~\ref{fig:RBdist}(a). In contrast, the spread for the CZ gate remains constant, see Fig.~\ref{fig:RBdist}(b). To investigate the reason for this observation we separately run different RB randomizations and plot individual histograms for each randomization in Fig.~\ref{fig:RBdist}(c). The individual histograms have an equal spread of $\mathcal{F}_{\rm seq}$, but their average sequence fidelity strongly depends on the chosen randomization. This indicates that the overall spread of $\mathcal{F}_{\rm seq}$ for the iSWAP gate is not given by statistical fluctuations, but result from a deterministic dependence of $\mathcal{F}_{\rm seq}$ on the individual RB sequences. This is a consequence of the presence of ZZ-type interactions that induce phase errors to qubits that depend on the state of other qubits.  Note that such state-dependent phase errors that are accumulated during a sequence influence the iSWAP gate because of its sensitivity to the alignment of the TC drive phase with the qubit's relative phase. The CZ gate does not depend on that phase, so the line shapes of the sequence fidelity histograms in Fig.~\ref{fig:RBdist}(b) remain small for increasing $m$.

\begin{figure}[!htbp]
\centering
\includegraphics[width =86mm]{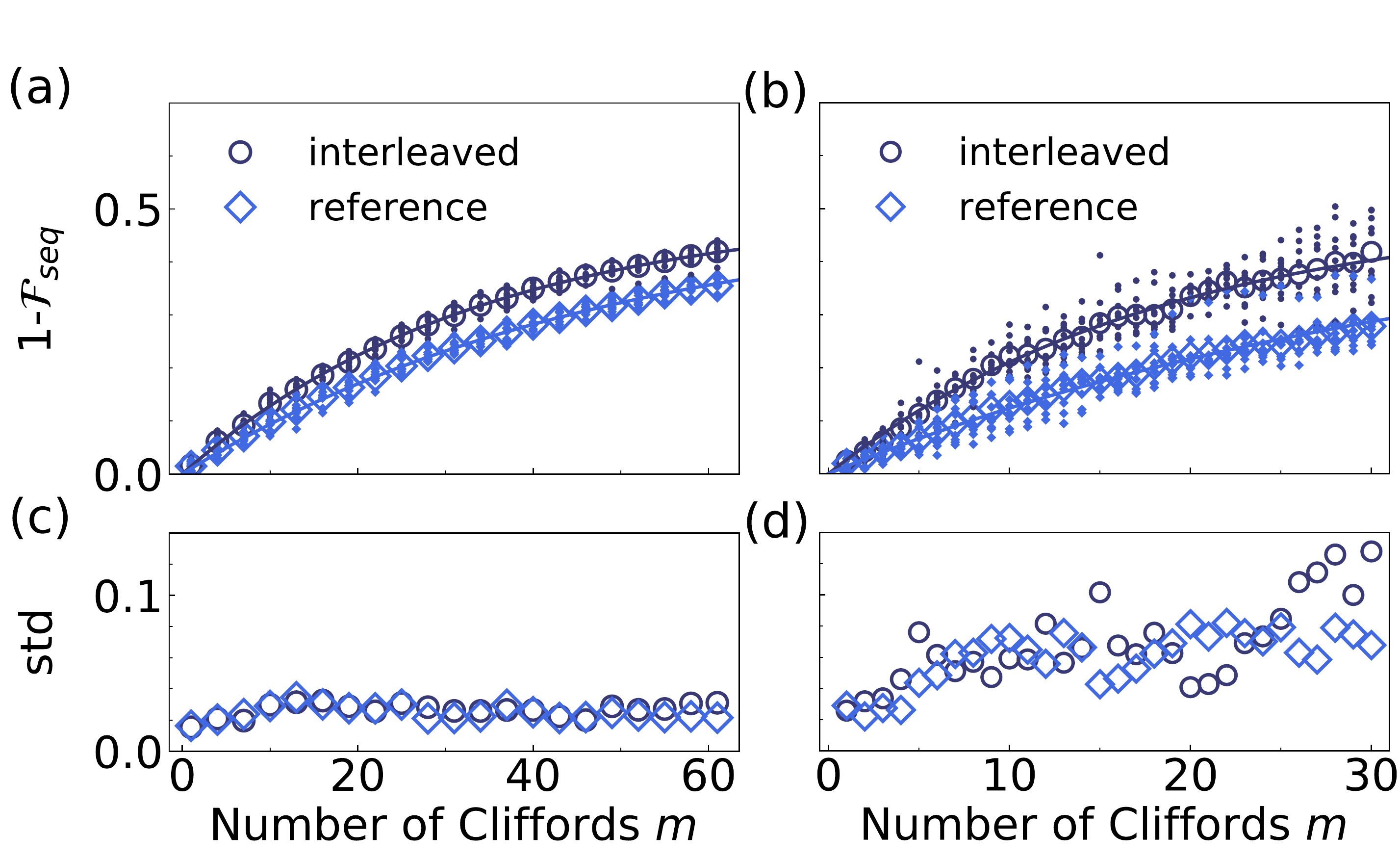}
\caption{RB simulations of (a),(c) CZ and (b), (d) iSWAP gates with $200~\rm{ns}$ length, using a simple two-level model of the two qubits and including dissipation and dephasing, as well as a static ZZ interaction of $-200~\rm{kHz}$. Shown in (a) and (b) are sequence fidelities of $10$ randomizations over different Clifford sequence lengths (small symbols) and their average (large symbol), as well as an exponential fit(solid lines) that determines the error per Clifford and thereby the error per gate ($0.0068$ for CZ,$0.0202$ for iSWAP). Due to the static ZZ crosstalk in the iSWAP gate, the RB fidelity at a given sequence length strongly depends on the randomization of the sequence, giving rise to large standard deviations of the state probability, as seen by comparing (c) and (d). \label{fig:RBsimulations}}
\end{figure}

In order to support this observation, we emulate the RB experiment using a simplified model where qubits are represented by two levels and a direct interaction between the two qubits is assumed, given by the unitaries of Eqs.~(\ref{eq:UCZphi}) and (\ref{eq:Uex}). A ZZ-type crosstalk  interaction (see Eq.~\ref{eq:heff_cphase}) with $\zeta/2\pi = -200~\rm{kHz}$ is added. The unitaries are transformed into Liouvillian representations of supermatrices, taking dissipation and decoherence into account, see \emph{Supplementary Material}. Ten different randomizations of Clifford sequences of variable lengths were simulated, and the resulting sequence fidelity is plotted in Fig.~\ref{fig:RBsimulations}. In agreement with the experimental observations, the spread in sequence fidelity increases for the iSWAP gate and stays much lower for the CZ gate. If no ZZ-type crosstalk is assumed, the spread in the iSWAP remains on the same level as that of the CZ gate, see \emph{Supplementary Material}. 

The misalignment between qubit phases and TC drive phase resulting from the ZZ-type crosstalk accumulates during a gate sequence. This also influences the RB error-per-gate that is larger than the one estimated from QPT. In numerical simulations (see \emph{Supplementary Material}) we find for the iSWAP gate with $\zeta/2\pi = -200~\rm{kHz}$ a RB error-per-gate of $0.0202$, whereas the QPT value is $0.0117$, almost a factor of two lower. If $\zeta = 0$ is chosen, the RB error decreases to $0.0066$ with a QPT error of $0.0078$.

\section{Conclusion}

On a given hardware, a 'native' gate set can be implemented that contains different variants of two-qubit gates. The preferred gate depending on the quantum algorithm that one wants to run, but also on the dominating source of noise that determines the fidelity of the gate operations. Here we demonstrate that in a tunable coupler architecture with a parametric gate implementation, the CZ gate is mostly insensitive to various error sources, such as ZZ-type crosstalk errors, frequency drifts of the qubits and phase errors of the drive. The error per gate of $0.89\%$ that is reached in our experiment for the CZ gate is mainly limited by decoherence of the qubits. In contrast, the error per gate of $1.3\%$ of the iSWAP gate is limited by external noise and ZZ-type interactions. 
Moreover, a large spread of the measured sequence fidelity is observed when sampling different realizations in a randomized benchmarking sequence. This puts the average error-per-gate as a measure for the quality of a gate in question, since the overall error after a sequence of gates depends significantly on the details of the sequence. Some quantum algorithms will therefore perform better, and some may lead to completely random results for the same number of gates.

The gate set in the tunable coupler architecture can be even further extended  by including 
the bSWAP gate, which requires modulation of the coupler at the qubits' sum frequency \cite{Roth2017}, typically around $10~\rm{GHz}$. Due to its dependence on the TC phase its noise sensitivity is expected to resemble the sensitivity of the iSWAP gate. A detailed analysis will, however, be subject to further investigation. 
 
To lower the gate errors for the iSWAP and bSWAP gate, improvements in the control electronics to reduce phase errors are required. A further route towards better gate fidelities  may be provided by optimizing the shape of the TC pulses to avoid errors such as leakage into the $|20\rangle$ state for the CZ gate on short time scales. On the device end, strategies to mitigate always-on ZZ-type interactions \cite{Mundada2019,Han2020} need to be devised that can be readily employed in scalable architectures. 

\section{Acknowledgments}

We thank the quantum team at IBM T. J. Watson Research Center, Yorktown Heights for insightful discussions and the provision of qubit devices. We thank R. Heller and H. Steinauer for technical support. This work was supported by the IARPA LogiQ program under contract W911NF-16-1-0114-FE, the ARO under contract W911NF-14-1-0124, the European Commission Marie Curie ETN project QuSCo (Grant Nr. 765267) and the European FET-OPEN project Quromorphic (Grant Nr.~828826).


%

\end{document}


\title{Benchmarking the noise sensitivity of different parametric two-qubit gates in superconducting qubits -- Supplementary Information}

\author{M. Ganzhorn}
\affiliation{IBM Research Zurich, S\"aumerstrasse 4, 8803 R\"uschlikon, Switzerland}
\author{G. Salis}
\affiliation{IBM Research Zurich, S\"aumerstrasse 4, 8803 R\"uschlikon, Switzerland}
\author{D.~J. Egger}
\affiliation{IBM Research Zurich, S\"aumerstrasse 4, 8803 R\"uschlikon, Switzerland}
\author{A. Fuhrer}
\affiliation{IBM Research Zurich, S\"aumerstrasse 4, 8803 R\"uschlikon, Switzerland}
\author{M. Mergenthaler}
\affiliation{IBM Research Zurich, S\"aumerstrasse 4, 8803 R\"uschlikon, Switzerland}
\author{C. M\"uller}
\affiliation{IBM Research Zurich, S\"aumerstrasse 4, 8803 R\"uschlikon, Switzerland}
\author{P. M\"uller}
\affiliation{IBM Research Zurich, S\"aumerstrasse 4, 8803 R\"uschlikon, Switzerland}
\author{S. Paredes}
\affiliation{IBM Research Zurich, S\"aumerstrasse 4, 8803 R\"uschlikon, Switzerland}
\author{M. Pechal}
\affiliation{IBM Research Zurich, S\"aumerstrasse 4, 8803 R\"uschlikon, Switzerland}
\author{M. Werninghaus}
\affiliation{IBM Research Zurich, S\"aumerstrasse 4, 8803 R\"uschlikon, Switzerland}
\author{S. Filipp}
\affiliation{IBM Research Zurich, S\"aumerstrasse 4, 8803 R\"uschlikon, Switzerland}

\maketitle

\section{Device characterization}
\label{sec:Device}

\begin{figure}[t]
\centering
\includegraphics[width = 0.95\textwidth]{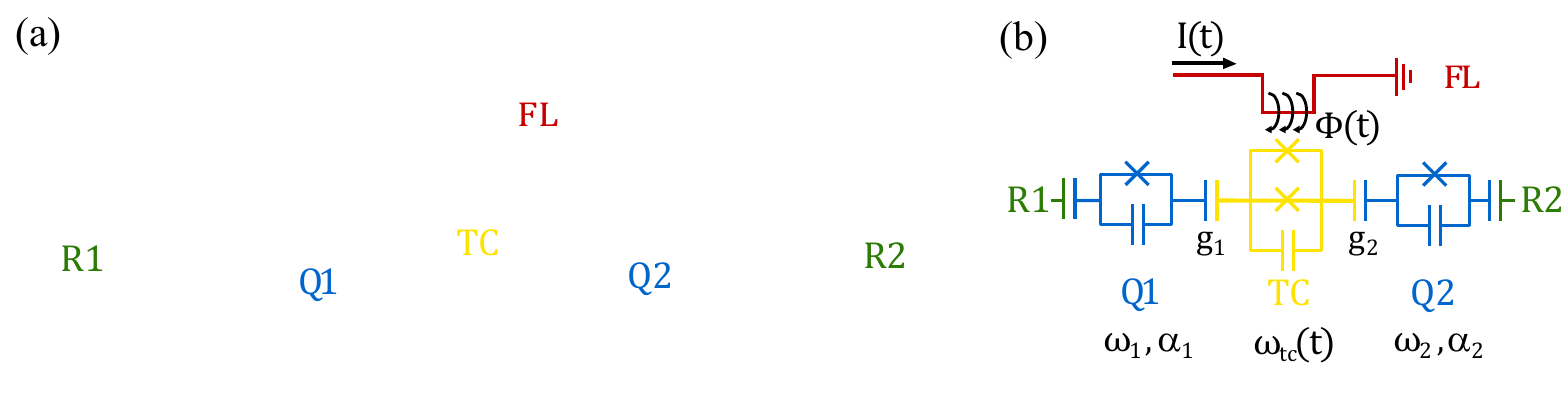}
\caption{(a) Optical micrograph and (b) circuit scheme of the device consisting of two fixed-frequency transmons (Q1, Q2) capacitatively coupled to a flux-tunable transmon acting as tunable coupler (TC). The tunable coupler is controlled by a flux line (FL) providing a current $I(t)$ and a consequent flux $\Phi(t) = \Phi_{\text DC} + \delta \cos(\ophi t + \phiphi)$ threading the SQUID-loop of the coupler. Each of the fixed-frequency qubits is coupled to an individual readout resonator (R1, R2).
\label{fig:chip}}
\end{figure}

Standard characterization techniques such as spectroscopy, Ramsey- and spin-echo experiments are used to extract the characteristic properties of the superconducting qubit device shown in Figure 
\ref{fig:chip}. The results are summarized in Table \ref{tab:params}.
\begin{table}[!htbp]
\centering
\resizebox{0.6\textwidth}{!}{%
\begin{tabular}{lcccccc}
\hline\hline
   & $\omega/2\pi$ (GHz) & $\alpha/2\pi$ (MHz)    		& $g/2\pi$ (MHz) 				& $T_1$ ($\mu$s) & $T_2$ ($\mu$s) & $T_2^*$ ($\mu$s) \\
Q1 & $5.089 $     			 & $-310 \pm 1$				  & $116 \pm 2$					  & $70 \pm 1$     & $81 \pm 1$     & $50 \pm 1$			 \\
Q2 & $6.189 $     			 & $-286 \pm 1$  				& $142 \pm 2$					  & $23 \pm 1$     & $26 \pm 1$     & $27 \pm 1$ 			 \\
TC & $8.1 $     			 & $-235 \pm 6$ 				& n/a        						& $15 \pm 1$     & $15 \pm 1$ 				& $7 \pm 0.1$ \\
\hline\hline
\end{tabular}}
\caption{Device parameters of two fixed-frequency transmons (Q1, Q2) coupled via a flux-tunable transmon (TC). The qubits have transition frequencies $\omega$, anharmonicities $\alpha$, and capacitive couplings $g$  to the tunable coupler (TC) at zero flux bias ($\Phi_{\text DC}=0$). The relaxation time $T_1$, spin-echo coherence time $T_2$ and Ramsey coherence time $T_2^*$ are measured at the flux bias point $\Phi_{\text DC}= 0.15\, \Phi_0$. 
\label{tab:params}}
\end{table}

\section{Flux-noise density of different tunable coupler designs}
 The geometry of the tunable coupler (TC) is similar to the one used in earlier experiments~\cite{Ganzhorn2019}, but for the different geometry of the SQUID loop. In these earlier experiments we have used symmetric SQUID loops with loop areas of $S= 625~\mu\rm{m}^2$. The gate fidelities and the accuracy of the reported quantum chemistry calculation in Ref.~\cite{Ganzhorn2019} were then mainly limited by the resulting flux noise amplitude $A = 17 \pm 0.4$ $\mu\Phi_0$ and the coherence time of the TC of $T_2^* = 30~\rm{ns}$. In order to improve the TC coherence time, we have characterized tunable couplers with different values for the SQUID asymmetry and different loop sizes. We have measured the $T_1$ and $T_2^*$ times as a function of the applied magnetic flux for the different TC designs using the technique described in~\cite{Ganzhorn2019} and have estimated the flux noise amplitude $A$ and density $A/S$ as summarized in Table \ref{tab:fluxnoise}. 

\begin{table}[h!]
\centering
\resizebox{0.4\textwidth}{!}{%
\begin{tabular}{lcccccc}
\hline\hline
Design  & $d$ 				& $S$ ($\mu$m$^2$)    		& $A$  ($\mu\Phi_0$) 				& $A/S$ (n$\Phi_0$/$\mu$m$^2$) 	\\
A 			& $0.59 $     & $225$			  						& $6.6 \pm 1.1$					  		& $29 \pm 5$     								\\
B 			& $0.36 $     & $225$			  						& $4.1\pm 1$					  		& $19 \pm 5$     								\\
C 			& $0.0 $     	& $100$ 									& $2.9 \pm 0.1$       								& $29 \pm 3$     								\\
D 			& $0.0 $     	& $625$ 									& $17.1 \pm 4$        								& $25 \pm 6$    								\\
\hline\hline
\end{tabular}}
\caption{
\label{tab:fluxnoise} Flux noise sensitivity for different SQUID loop areas $S$ and asymmetries $d$ measured by the flux noise amplitude $A$ assuming $1/f$ noise with a power spectral density of $A^2/\omega$.}
\end{table}

\begin{center}
\begin{figure}[h!]
\includegraphics[width=0.8\textwidth]{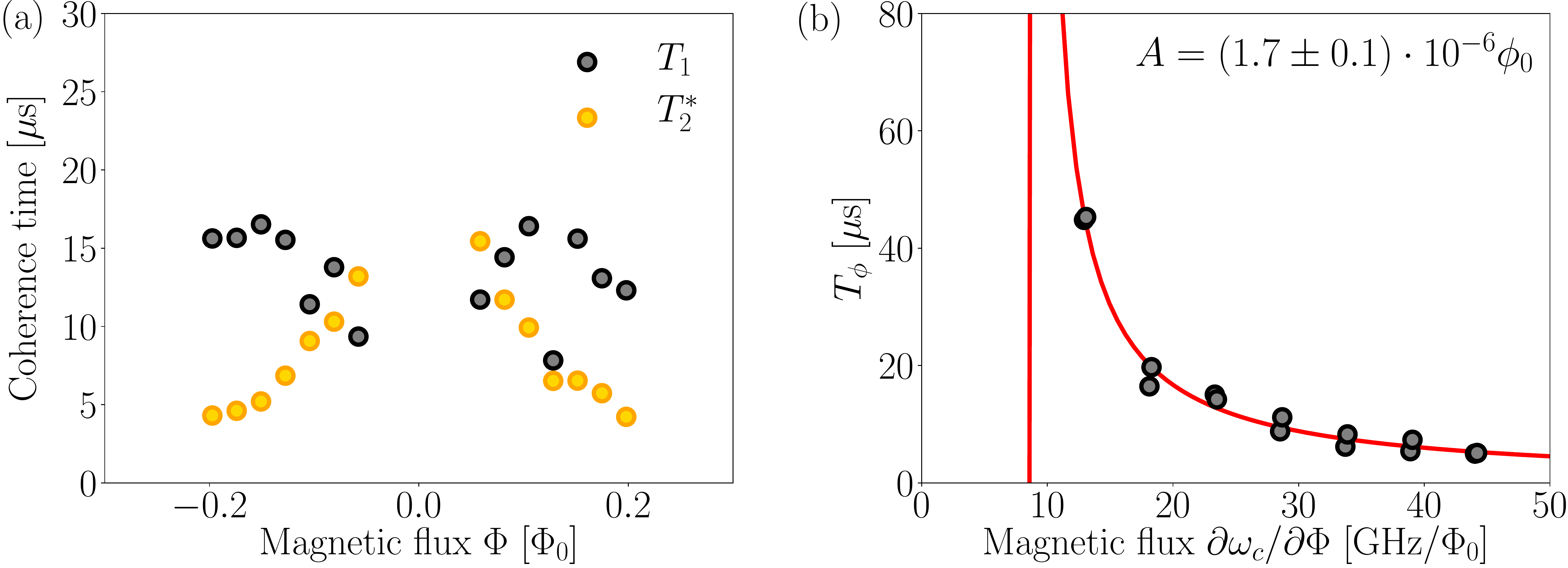}
\caption{Coherence of the tunable coupler. (a) Measurement of $T_1$ and $T_2^*$ time as function of magnetic flux. (b) $T_{\phi}$ as function of $\partial \omega_c / \partial \Phi$. The flux noise amplitude $A$ is obtained from an exponential fit (red line). \label{fig:coherence_tc}}
\end{figure}
\end{center}

While the TC asymmetry has only a minor effect on the flux-noise amplitude, a reduction of the loop size to $100~\mu\rm{m}^2$ as in design C reduces the noise sensitivity. The small loop size, however, has weaker inductive coupling to the drive line and  requires a larger current. This leads to heating of the device. We have, therefore, used the TC design B with a medium-sized SQUID loop ($S=225~\mu\rm{m}$) for the experiments reported in this manuscript. Later, we have improved the filtering of the DC and RF lines used for the parametric drive and further reduced the flux noise amplitude to $A = 1.7 \pm 0.1$ $\mu\Phi_0$ (Fig.\ref{fig:coherence_tc}) for the  experiments.

\section{Calibration}

\subsection{CZ-gate calibration}
In the following, we describe the complete calibration procedure for a controlled phase gate $\rm{CZ}_\varphi$ described by the unitary
\begin{align}
\label{eq:UCZphi}
U_{\rm CZ}(\varphi) & = 
\begin{pmatrix} 
1 & 0 & 0 & 0 \\
0 & 1 & 0 & 0 \\ 
0 & 0 & 1 & 0 \\ 
0 & 0 & 0 & e^{-i\varphi} \\ 
\end{pmatrix}.
\end{align}
Setting the phase $\varphi$ to $\pi$ realizes a controlled-Z (CZ) gate.
A $\rm CZ_{\varphi}$-type gate with  arbitrary phase $\varphi$ is implemented in the tunable coupler architecture using higher transmon levels. Threading a magnetic flux $\Phi(t) = \Phi_{\text{DC}} + \delta_{\Phi} \cos(\omega_\Phi t+\eta)$ at a frequency that is slightly detuned by $\Delta = \omega_{\Phi} - \omega_\alpha$ from the $\ket{11}-\ket{20}$ transition for a time $\tau_{2 \pi} = 2 \pi/\sqrt{\Omega_{\rm eff}^2+\Delta^2}$ through the SQUID loop of the TC implements the following unitary transformation with a geometric phase $\phi_g$~\cite{Krantz2019}

\begin{align}
\label{eq:Uczuncal}
U_{\rm CZ_{\phi_g}} & = 
\begin{pmatrix} 
1 & 0 & 0 & 0 \\
0 & e^{-i\phi_{01}} & 0 & 0 \\ 
0 & 0 & e^{-i\phi_{10}} & 0 \\ 
0 & 0 & 0 & e^{-i(\phi_g+\phi_{\zeta}+\phi_{10}+\phi_{01})} \\ 
\end{pmatrix}.
\end{align}
Here we assume that the system returns to the $\ket{11}$ state and leakage to the $\ket{20}$ can therefore be neglected.
$\phi_{\zeta}$ is the phase shift induced by the static and induced ZZ-type coupling between the qubits discussed in Section \ref{sec:zz_shift} below, and the single qubit phases $\phi_{01}$ and $\phi_{10}$ are caused by dispersive shifts from the TC on the qubits during the gate. $\Omega_{\rm eff} = \Omega_{\rm eff}(\Phi_{\text{DC}},\delta_{\Phi})$ denotes the Rabi rate of the $\ket{11}-\ket{20}$ transition. For the pulses we use a Gaussian flat-top  with a rise-/falltime of $\sim13~\rm{ns}$. In contrast to exchange-type (iSWAP) gates discussed below, the geometric phase $\phi_g$ of such a $\rm CZ_{\phi_g}$-type gate does not depend on the phase $\varphi_{\Phi}$ of the two-qubit gate drive and therefore the CZ gate is not sensitive to type B errors. 

\begin{figure*}[hbt!]
\centering
\includegraphics[width = 0.98\textwidth]{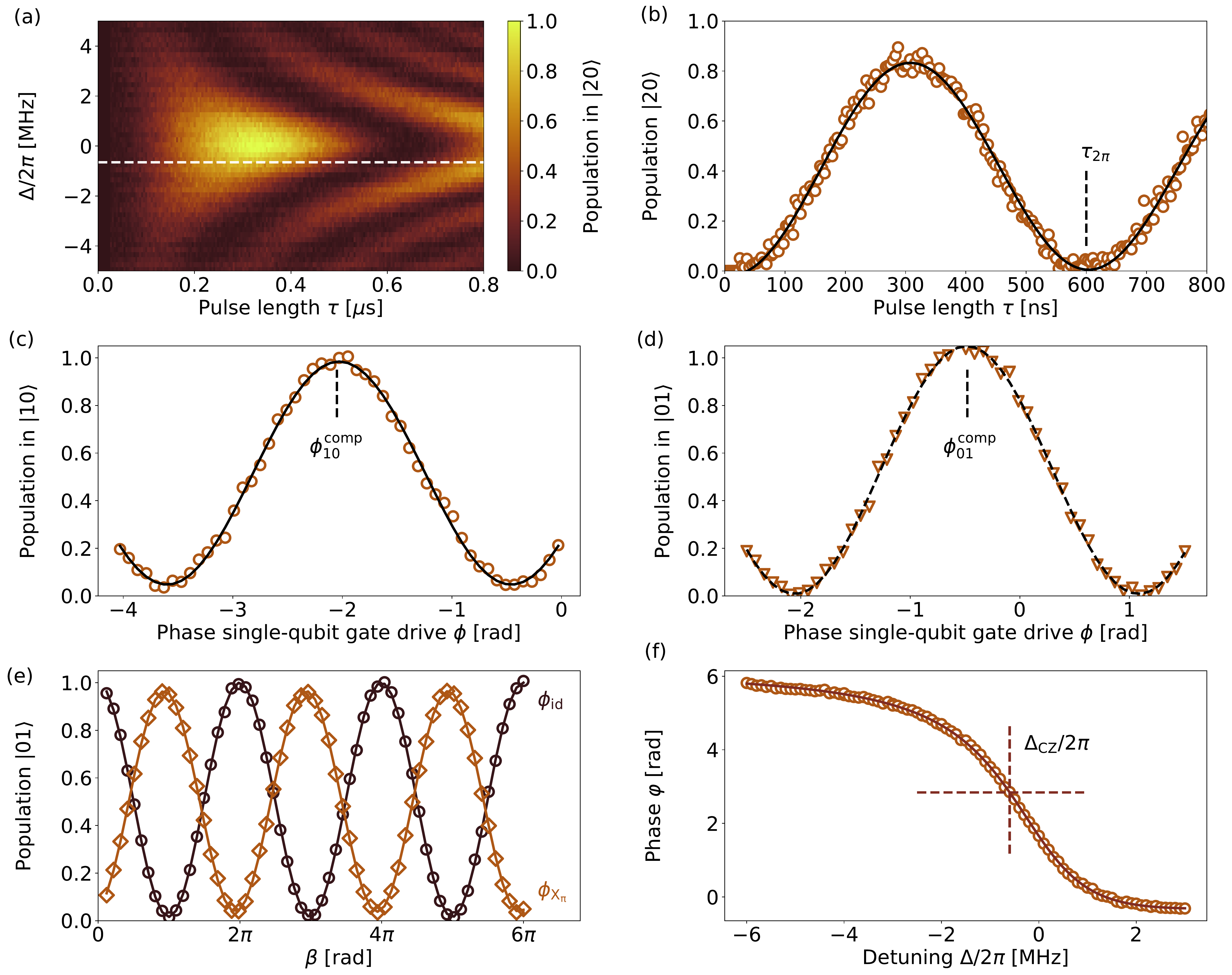}
\caption{Calibration of the CZ-gate $U_{\rm CZ_{\pi}}$ at a TC flux-bias point $\Phi_{\rm DC} = 0.15 \Phi_0$: 
(a) Population in $\ket{20}$ as function of pulse length $\tau$ and detuning from the $\ket{11}-\ket{20}$ transition $\Delta$. 
(b) Population in $\ket{20}$ as function of pulse length $\tau$ at a given drive frequency $\omega_{\Phi}$ (white dashed line in (a)). For a pulse length of $\tau_{2\pi}$ the population is returned to the initial $\ket{11}$ state. 
(c) Population in $\ket{10}$ as function of the phase $\phi$ of the a virtual $Z_{\phi}$-gate of qubit $Q_1$ (open circles). Solid line is a sinusoidal fit used to determine the compensation phase $\phi_{10}^{\rm comp}$. 
(d) Population in $\ket{01}$ as function of the phase $\phi$ of the a virtual $Z_{\phi}$-gate of qubit $Q_2$ (triangles). Dashed line is a sinusoidal fit to determine the compensation phase  $\phi_{01}^{\rm comp}$. 
(e) Population in $\ket{01}$ measured for different rotation angles $\beta$ using the calibration sequence in Eq. (\ref{eq:cal_phi_cz}) with $U = I$ (yellow curve) and $U = X_{\pi}$ (brown curve). Solid lines represent fits with Eq. (\ref{eq:id}) and (\ref{eq:cz}), respectively. 
(f) Phase $\varphi$ as function of the detuning $\Delta$. 
\label{fig:gate_cal_cz}}
\end{figure*}

To calibrate the gate, we first measure Rabi oscillations (starting in the $\ket{11}$ state) as a function of the detuning $\Delta$ as shown in Fig.~\ref{fig:gate_cal_cz}(a). For each detuning, this determines the gate length $\tau_{2\pi}$ at which the population completely returns to the $\ket{11}$ state.  Fig.~\ref{fig:gate_cal_cz}(b) shows the oscillations for a detuning of $\Delta/2\pi = -650~\rm{kHz}$. Next, we compensate single-qubit phase shifts $\phi_{01}$ and $\phi_{10}$ (type D errors). Starting with the $\ket{00}$ state, one of the two qubits is rotated by $\pi/2$ about the $X$-axis (denoted as  $X_{\frac{\pi}{2}}$ pulse). Applying the $\rm CZ_{\phi_g}$-gate and its inverse should yield the identity operation on that qubit, such that a final $X_{\frac{\pi}{2}}$ pulse will lead to population transfer into the excited state. Adding an extra $Z$-rotation after the $\rm CZ_{\phi_g}$ gate as in the circuit 

\begin{equation}
\begin{array}{c}
\Qcircuit @C=0.75em @R=0.5em {
& Q1 & \quad & \ket{0} & \quad & \gate{X_{\frac{\pi}{2}}} & \multigate{1}{U_{\rm CZ_{\phi_g}}} & \gate{Z_{\phi}} & \multigate{1}{U_{\rm CZ_{\phi_g}}^\dag} & \gate{Z_{\phi}} & \gate{X_{\frac{\pi}{2}}} \\
& Q2 & \quad & \ket{0} & \quad & \qw & \ghost{U_{\rm CZ_{\phi_g}}} & \qw & \ghost{U_{\rm CZ_{\phi_g}}^\dag} & \qw & \qw &
}
\end{array}
\label{eq:cal_sqphi_cphase}
\end{equation}

with phase $\phi$ allows for the adjustment of the single qubit phase shift. $Z_{\phi}$ is a virtual-Z gate \cite{McKay2017} with programmable phase $\phi$ defined by
\begin{align}
\label{Zgate}
Z_{\phi} = 
\begin{pmatrix} 
e^{-i\phi/2} & 0 \\
0 & e^{i\phi/2}\\ 
\end{pmatrix}
\end{align}
In practice, the $Z_{\phi}$ gate is realized by shifting the phase of the corresponding single qubit drive signal by $\phi$.  The resulting oscillation of the qubit $Q_1$ population is shown in Fig.~\ref{fig:gate_cal_cz}(c). The same sequence with now applying the $X_{\frac{\pi}{2}}$ pulse on $Q_2$ allows for compensation of the $\phi_{01}$ single-qubit phase shift.
The resulting unitary transformation reads 
\begin{align}
\label{Uczfinal}
U_{\rm CZ_{\varphi}} = Z_{\phi_{10}}^1\cdot Z^2_{\phi_{01}}\cdot U_{\rm CZ_{\phi_g}}& = 
\begin{pmatrix} 
1 & 0 & 0 & 0 \\
0 & 1 & 0 & 0 \\ 
0 & 0 & 1 & 0 \\ 
0 & 0 & 0 & e^{-i\varphi}\\ 
\end{pmatrix}.
\end{align}
The total phase $\varphi = \phi_g + \phi_\zeta$ includes the geometric phase $\phi_g$ and the extra phase $\phi_\zeta = \zeta_{\rm stat} \tau$ from the ZZ-type coupling between the qubits.
As discussed in the main text, the geometric phase
\begin{equation}
\varphi_g = \pi \left\{1 - \cos \left( \arctan \left[\frac{\Omega_{\rm eff}}{\Delta}\right]\right)\right\}.
\label{eq:geom}
\end{equation}
originates from the cyclic evolution of the state $\ket{11}$.

As the last step, we adjust the \emph{total} phase to give $\varphi=\pi$, which also compensates static ZZ-shifts during the gate operation. For that purpose we run the  $\pi$/no-$\pi$ sequence 
\begin{equation}
\begin{array}{c}
\Qcircuit @C=1em @R=0.5em {
& Q1 & \quad & \ket{0} & \quad & \gate{U} & \multigate{1}{U_{\rm CZ_{\varphi}}} & \qw & \qw &\\
& Q2 & \quad & \ket{0} & \quad & \gate{X_{\frac{\pi}{2}}} & \ghost{U_{\rm CZ_{\varphi}}} & \gate{\tilde{X}_{\frac{\pi}{2}}} & \qw &
}
\end{array}
\label{eq:cal_phi_cz},
\end{equation}
with either a leading identity $U = I$  or a $\pi$-rotation $U = X_{\pi}$. Here, $\tilde{X}_{\frac{\pi}{2}} = \cos\beta X + \sin\beta Y$ described a $\pi/2$ pulse about the  axis rotated by $\beta$ in the azimuthal $X-Y$ plane. The population measured on qubit $Q_2$ is described by 
\begin{eqnarray}
p(\beta) &=& \frac{1}{2} + \frac{1}{2} \cos{\beta} \quad \text{for} \quad U = I \label{eq:id}\\
p(\beta) &=& \frac{1}{2} + \frac{1}{2} \cos{\left( \beta+\varphi\right)} \quad \text{for} \quad U = X_{\pi} \label{eq:cz}
\end{eqnarray}

Measuring the population in qubit $Q_2$ as a function of $\beta$ for both cases as shown in Fig.~\ref{fig:gate_cal_cz}(e) [Fig.~2(b) in the main text] and fitting the data with Eq.~(\ref{eq:id}) and (\ref{eq:cz}) yields the controlled phase shift on the $\ket{11}$ state as the sum of geometric phase and static offset phase $\varphi = \varphi_g + \phi_{\zeta}$ for a given detuning $\Delta$. The procedure is repeated for different detuning values, the resulting phase $\varphi$ is shown in Fig.~\ref{fig:gate_cal_cz}(f) [Fig.~2(c) of the main text]. The targeted CZ-gate with $\varphi = \pi$ is realized at the detuning $\Delta_{\rm CZ}=-770~\rm{kHz}$.

\begin{figure*}[htb!]
\centering
\includegraphics[width = 0.45\textwidth]{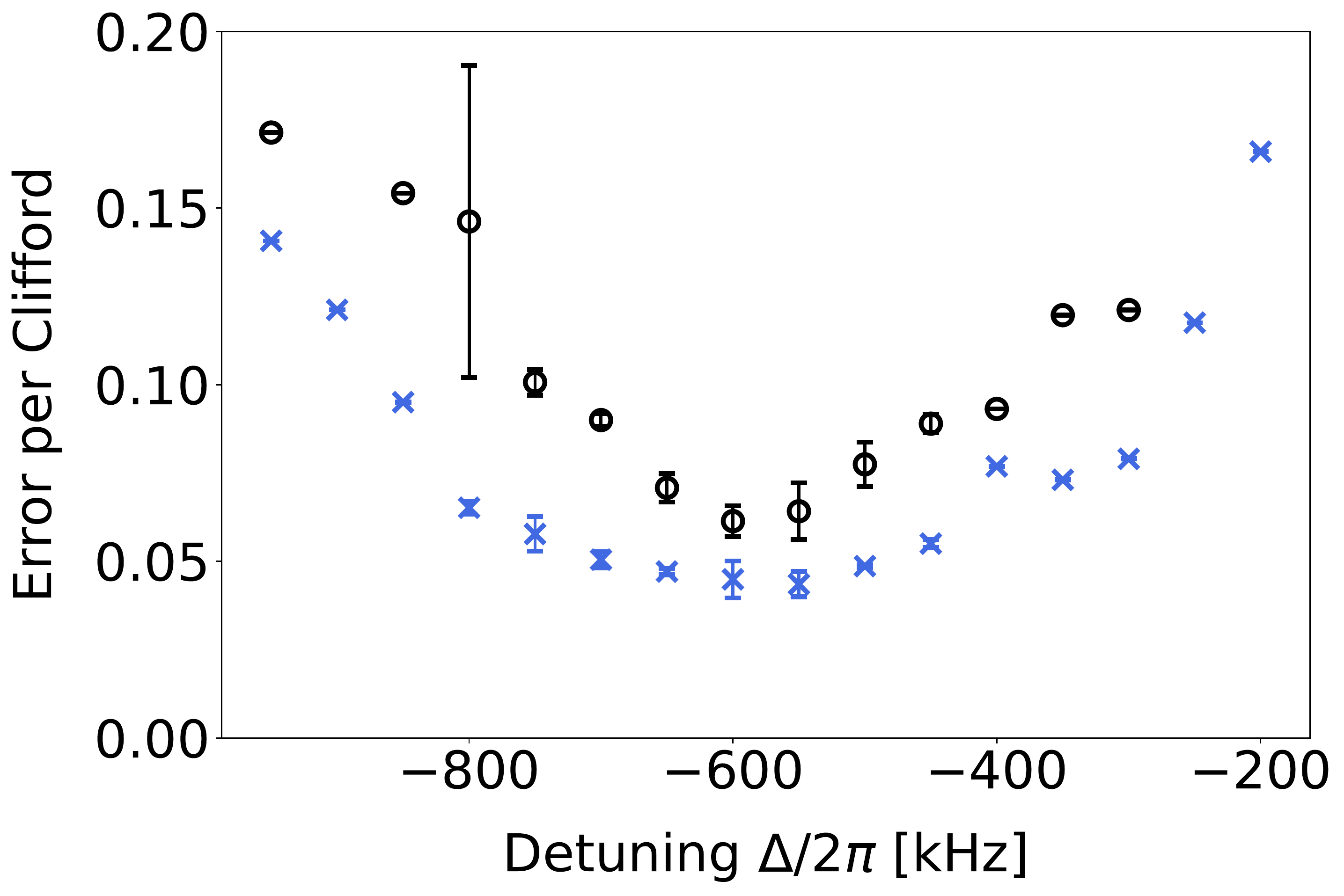}
\caption{Error per Clifford as a function of the drive detuning $\Delta$. Shown are the interleaved (black circles) and the reference (blue crosses) randomized benchmarking curves.\label{fig:detuningRB}}
\end{figure*}

To fine tune $\Delta_{\rm CZ}$ we characterize the error per CZ gate using interleaved RB as a function of detuning shown in Fig.~\ref{fig:detuningRB}. We find the minimum of the error at a slightly different value $\Delta_{\rm CZ}=-600~\rm{kHz}$. Possible explanations of this mismatch between the value found in calibration sequences and the value found by fine-tuning with randomized benchmarking experiments are averaging effects of single-qubit phases in the RB sequence, pulse distortions and/or memory effects caused by reflections in the system.

\subsection{iSWAP-gate calibration}
\label{sec:iswap_cal}

The interaction Hamiltonian $H_{\rm iSWAP}(\Omega_{\rm eff},\varphi)$ [main text Eq.~(6)] and the corresponding unitary operator 
\begin{align}
\label{eq:Uex}
U_\text{iSWAP}(\theta,\varphi) =\begin{pmatrix} 
1 & 0 & 0 & 0 \\
0 & \cos\theta/2 & ie^{i\varphi}\sin\theta/2 & 0 \\ 
0 & ie^{-i\varphi}\sin\theta/2 & \cos\theta/2 & 0 \\ 
0 & 0 & 0 & 1 \\ 
\end{pmatrix}
\end{align}
result in the iSWAP gate when setting the angle $\theta=\pi$.
 $\theta = \Omega_{\rm eff} \tau$ is controlled by the length $\tau$ of the parametric drive pulse on the tunable coupler. For a length $\tau_{\pi}=\pi/\Omega_{\rm eff}$ an iSWAP gate is realized, which completely transfers an excitation from one qubit to the other. 

In the following, we will limit our discussion to the iSWAP gate $U_{\rm iSWAP_{\varphi}}=U_\text{iSWAP}(\pi,\varphi)$, a Clifford-gate whose error can be estimated via randomized benchmarking.

\begin{figure*}[htb!]
\centering
\includegraphics[width = 0.98\textwidth]{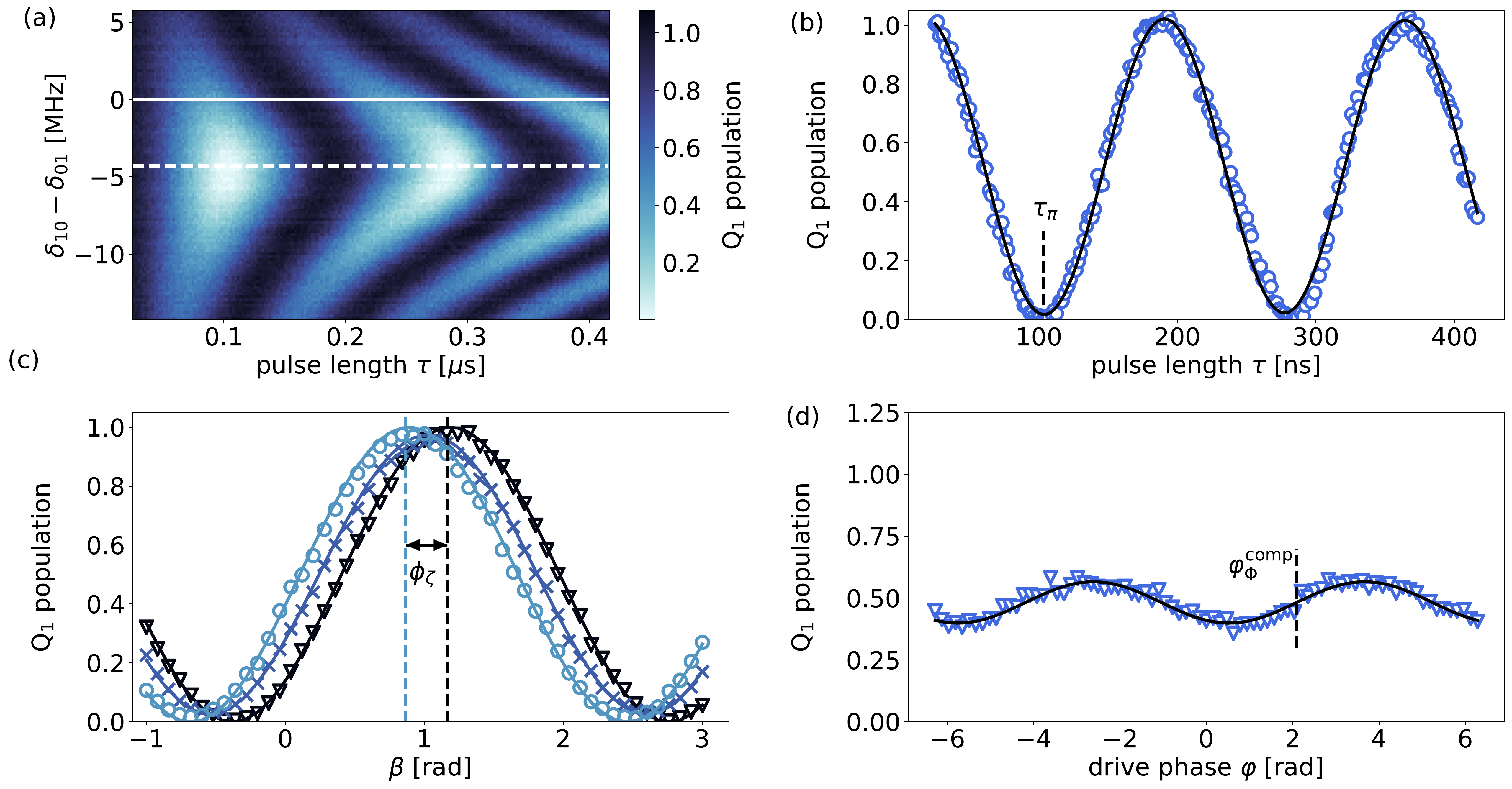}
\caption{Calibration routine for an iSWAP gate $U_{\rm iSWAP(\pi,0)}$ at $\Phi_{\rm DC} = 0.15 \Phi_0$: (a) Population in qubit $\ket{10}$ as function of pulse length $\tau$ and the detuning $(\delta_{01}-\delta_{10})$. The white dashed (solid) line indicates the driven (undriven) resonance frequency. (b) Population in $\ket{10}$ as function of pulse length $\tau$ at the driven resonance frequency (white dashed line in (a)). For a pulse length of $\tau_{\pi} = 103$ ns an iSWAP gate $U_{\rm iSWAP(\pi,\varphi)}$ is realized. (c) Population in $\ket{10}$ as function of the phase $\phi$ obtained from sequence \ref{eq:cal_sqphi_swap} with $U=I$ (black symbols), $U=X_{\pi/2}$ (blue symbols) and $U=X_{\pi}$ (light blue symbols). Solid lines are fits to the functions discussed in the text. (d) Population in $\ket{01}$ as function of the phase $\varphi_{\Phi}$ of the coherent drive. Black solid line is a fit with Eq.~(\ref{phi_phase_cal}). \label{fig:gate_cal}}
\end{figure*}

For the calibration of the iSWAP gate we first determine the frequency and pulse area of the flux pulse implementing the transformation $U_{\rm iSWAP(\pi,\varphi)}$. As for the CZ gate we use a Gaussian flat-top  with a rise-/falltime of $\sim13~\rm{ns}$. The shift of the mean TC frequency  causes dispersive shifts $\delta_{01}$ and $\delta_{10}$ of the qubit frequencies during the TC pulse, which also shifts the iSWAP transition frequency by $\delta_{01}-\delta_{10}$. 
The resonance drive frequency, thus, depends on the drive amplitude~\cite{Roth2017}. Consequently, we fix the pulse amplitude and measure Rabi oscillations for frequencies at a detuning from the qubits' difference frequency as shown in Fig.~\ref{fig:gate_cal}(a). For a given pulse modulation amplitude, we select the frequency with maximum contrast oscillations (minimal Rabi frequency) and set the gate length $\tau_{\pi}$ for which $\theta=\pi$, see Fig.~\ref{fig:gate_cal}(b). 

The dispersive shifts $\delta_{i}$ also induce corresponding phase shifts $2 \pi \delta_{01/10} \tau$ on the states $\ket{01}/\ket{10}$, which requires an adjustment of the qubits' reference frame after the gate operation. 
The precise value of the Z-rotations on the individual qubits is determined by a Ramsey-type sequence. For qubit $Q_1$ the sequence
\begin{equation}
\begin{array}{c}
\Qcircuit @C=0.75em @R=0.5em {
& Q1 & \quad & \ket{0} & \quad & \gate{X_{\frac{\pi}{2}}} & \multigate{1}{U_{\rm iSWAP}} & \multigate{1}{U_{\rm -iSWAP}} & \gate{\tilde{X}_{\frac{\pi}{2}}} \\
& Q2 & \quad & \ket{0} & \quad & \gate{U} & \ghost{U_{\rm iSWAP}}  & \ghost{U_{\rm -iSWAP}}  & \qw &
}
\end{array}
\label{eq:cal_sqphi_swap}
\end{equation}
with the rotated trailing $\pi/2$-pulse about the $\tilde{X} = \cos(2\beta) X + \sin(2\beta) Y$ axis and with $U=I$
results in an oscillation of the population in the $\ket{10}$-state as a function of the variable phase $\beta$, $p(\beta) = \{1 + \cos\left[2(\beta+\phi_{10})\right]\}/2$, see Fig.~\ref{fig:gate_cal}(c). This fit function is found by assuming that the shift of the transition frequency is already compensated by the drive frequency and that the length of the gate $U_{\rm iSWAP}$ is adjusted to give a complete excitation transfer. For $\beta=-\phi_{10}$ the excitation probability of $Q_1$ is maximal, and $U_{\rm iSWAP}U_{\rm -iSWAP}$ equals the identity operation. Similarly, the phase shift on qubit $Q_2$ is determined by exchanging the roles of $Q_1$ and $Q_2$. After each iSWAP gate, the phase shifts $\phi_{01}$ and $\phi_{10}$ are then compensated by additional virtual-Z gates on each qubit, as is the phase $\varphi_\Phi$ of the coupler drive by an amount $\Delta\varphi_{\Phi,0}=\phi_{01}-\phi_{10}$.

The situation is complicated by the presence of the ZZ-type interactions during the gate operation described by the Hamiltonian \cite{Krantz2019}
\begin{eqnarray}
H_\zeta/\hbar = \zeta \ket{11}\bra{11} = \frac{\zeta}{4} \left(ZZ-IZ-ZI+II \right),
\label{eq:heff_cphase}
\end{eqnarray}
where $\zeta = \omega_{11} - \omega_{01} - \omega_{10}$ is the energy shift of the $\ket{11}$ state.  
While the iSWAP gate acts only on the $\ket{01}-\ket{10}$ subspace, the $\ket{11}$ state acquires a phase  $\phi_{\zeta}$ proportional to $\zeta \tau$ relative to the sum of the individual qubit phases. With the calibration sequence described above, only the dispersive phase shifts on the states $\ket{01}$ and $\ket{10}$ are zeroed. When the system is initially in the $\ket{11}$ state a ZZ-type error $\ket{11}\rightarrow \exp{-i\zeta \tau}\ket{11}$ is picked up. Alternatively, in a calibration with $Q_2$ in the excited state ($U=X_{\pi}$) the Ramsey-oscillation are shifted by the ZZ-shift $\phi_\zeta$ according to $p(\beta) = \{1 + \cos\left[2(\beta+\phi_{10}+\phi_\zeta)\right]\}/2$  [Fig.~\ref{fig:gate_cal}(c); black triangles] resulting in $\beta=-\phi_\zeta -\phi_{10}$. With this choice, however, the system accumulates a phase error whenever it is in the ground state.  From randomized benchmarking experiments we have identified that balancing the $\phi_\zeta$ error results in the lowest errors per gate: 
With qubit $Q_2$ prepared in a superposition state ($U=X_{\pi/2}$) the Ramsey-oscillations are shifted by $\phi_\zeta/2$ following $p(\phi) = \{1 +  \cos{\phi_{\zeta}}\cos{ \left[ 2(\beta+\phi_{10})+\phi_{\zeta}\right]}/2$  resulting in $\beta = -\phi_\zeta/2 -\phi_{10}$ [Fig. \ref{fig:gate_cal}(c); blue crosses)]. A virtual-Z shift of value $\beta$ compensates the dispersive shift on qubit $Q_1$ up for a phase factor $\phi_\zeta/2$. A similar measurement with the roles of $Q_1$ and $Q_2$ exchanged determines the virtual-Z shift for qubit $Q_2$.  A fit of the measurement to the indicated curves gives a phase shift $\phi_{\zeta} = -0.299~\rm{rad}$ which corresponds to a shift of the $\ket{11}$ state of $\zeta = -464~\rm{kHz}$ for the calibrated pulse length of $\tau_{\pi} = 103~\rm{ns}$. The shift of the $\ket{11}$ state has a contribution from a static coupling of $\zeta_{\rm stat} = -202~\rm{kHz}$ and a remaining contribution from a drive induced coupling of $\zeta_{\rm dyn} = \zeta - \zeta_{\rm stat} = -268~\rm{kHz}$, as discussed in the following Section~\ref{sec:zz_shift}. 

With the such compensated single-qubit dispersive shifts we obtain the gate operation
\begin{align}
\label{Uexp_comp}
\nonumber \tilde{U}_{\rm iSWAP_{\varphi}} & = Z_{\phi_{01}^{\rm comp}} \cdot Z_{\phi_{10}^{\rm comp}} \cdot U_{\rm iSWAP_{\varphi}} \\
& = 
\begin{pmatrix} 
1 & 0 & 0 & 0 \\
0 & 0 & -i e^{i(\varphi-\phi_{\zeta}/2)} & 0 \\ 
0 & -i e^{i(-\varphi-\phi_{\zeta}/2)} & 0 & 0 \\ 
0 & 0 & 0 & 1 \\ 
\end{pmatrix}.
\end{align}
The phase $\varphi=\varphi_\Phi +\varphi_0$ is set to zero by adjusting the phase $\varphi_{\Phi}$ of the parametric drive to compensate an a priori unknown offset phase $\varphi_0$. We use the  cross-Ramsey pulse sequence
\begin{equation}
\begin{array}{c}
\Qcircuit @C=1em @R=0.5em {
& Q1 & \quad & \ket{0} & \quad & \gate{X_{\frac{\pi}{2}}} & \multigate{1}{\tilde{U}_{\rm iSWAP_{\varphi}}} & \gate{X_{\frac{\pi}{2}}} & \qw &\\
& Q2 & \quad & \ket{0} & \quad & \gate{X_{\frac{\pi}{2}}} & \ghost{\tilde{U}_{\rm iSWAP_{\varphi}}} & \gate{Y_{\frac{\pi}{2}}} & \qw &
}
\end{array}
\label{eq:cal_phi_swap}
\end{equation}
with a leading $X_{\frac{\pi}{2}}$ pulse on $Q_2$ which prepares a superposition state to average over the ZZ-shifts. 

The population measured in qubit $Q_2$ depends on the phase $\varphi_\Phi$ [Fig.~\ref{fig:gate_cal}(d)] and is described by the following equation
\begin{eqnarray}
p(\varphi_{\Phi}) = \frac{1}{2}\big\{1-\sin{(\varphi_\Phi+ \varphi_0)} \sin{(\phi_{\zeta}/2)}\big\}
\label{phi_phase_cal}
\end{eqnarray}
From $p(\varphi_\Phi)=1/2$ we find the value for the drive phase $\varphi_{\Phi} = -\varphi_0$ and realize the iSWAP operation up to ZZ-induced phase shifts, 
\begin{align}
\label{Uexp_final}
U_{\rm iSWAP} & = 
\begin{pmatrix} 
1 & 0 & 0 & 0 \\
0 & 0 & -ie^{-i\phi_{\zeta}/2} & 0 \\
0 & -ie^{-i\phi_{\zeta}/2} & 0 & 0 \\
0 & 0 & 0 & 1 \\ 
\end{pmatrix}.
\end{align}
Note that for small ZZ-shift two separate measurements with a leading identity or $\pi$-pulse on $Q_2$ are employed to find the correct value for $\varphi_\Phi$.

Finally, we fine tune the frame correction $\Delta\varphi_{\Phi}$ of the TC drive caused by the dispersive qubit shifts. We characterize the error per iSWAP gate using interleaved RB as a function of $\Delta\varphi_\Phi$ as shown in Fig.~\ref{fig:detuningRB}. We find the minimum of the error very close to $\Delta\varphi_{\Phi,0}$. Small asymmetries may be due to a slightly miscalibrated absolute phase $\eta$ to which the repetitive influence of the frame correction $\Delta\varphi_\Phi$ is more sensitive. Such effects are confirmed in numerical simulations.  

\begin{figure*}[!htbp]
\centering
\includegraphics[width = 0.45\textwidth]{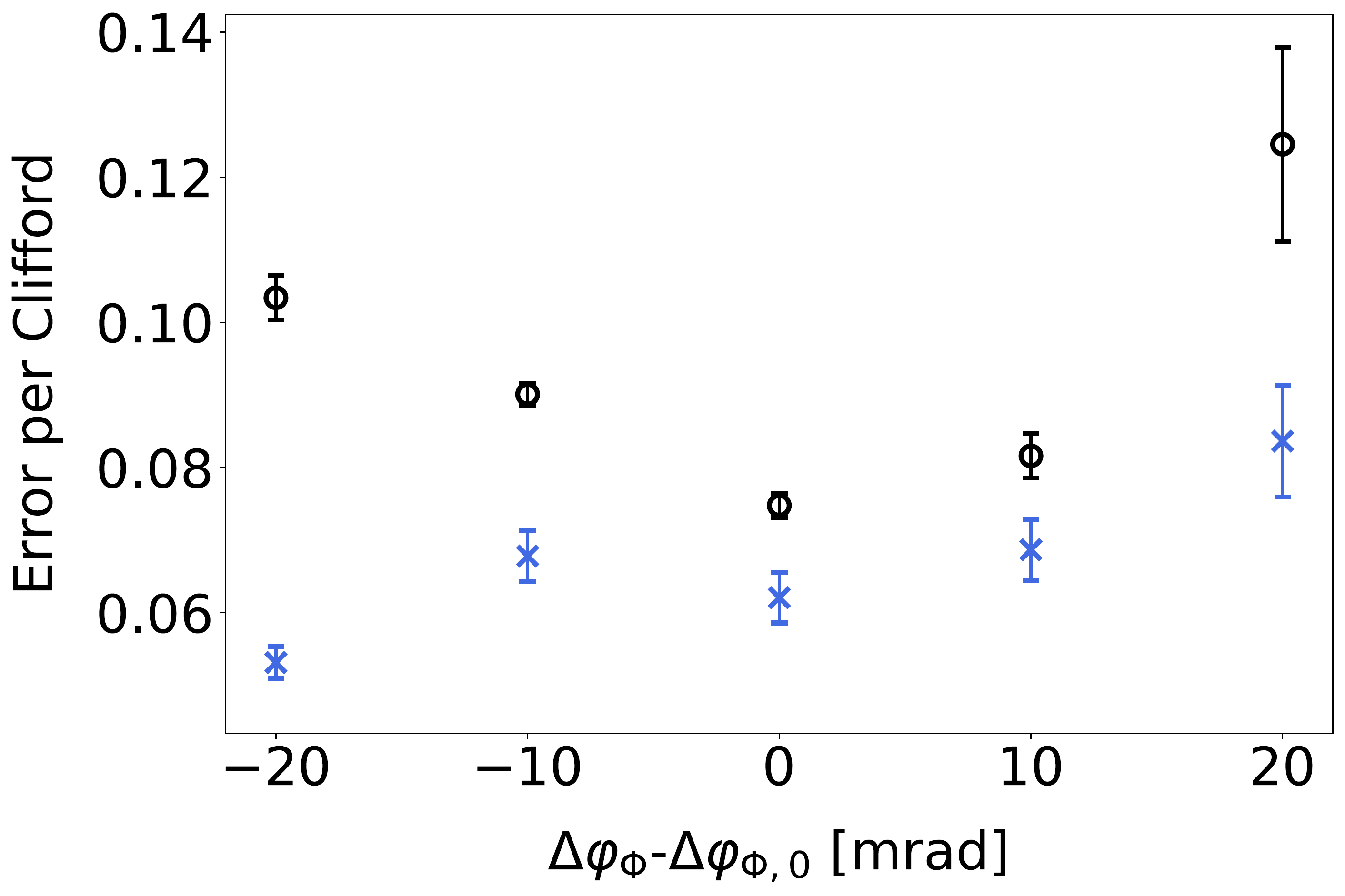}
\caption{Error per Clifford for varying TC frame correction phases  $\Delta\varphi_{\Phi}$, measured for a gate length of 171 ns. Shown are the interleaved (black circles) and the reference (blue crosses) randomized benchmarking curves. The phase is shown relative to the calibrated value $\Delta\varphi_{\Phi,0}=\phi_{01}-\phi_{10}$  \label{fig:varphiRB}}
\end{figure*}

\section{ZZ-type dispersive shifts}
\label{sec:zz_shift}
With the weakly anharmonic structure of the transmon, the second-excited state of the transmon influences the energy levels of the system. One of the effects is the dispersive shift of the $\ket{11}$ due to its coupling to the $\ket{20}$ and $\ket{02}$ level. In addition to the static dispersive shift, we also consider drive-induced ZZ-type interactions caused by the mean frequency shift of the tunable coupler during the gate operation.

\subsection{Static ZZ-shift}

The static ZZ shift $\zeta_{\rm stat}$ is mediated by the tunable coupler and depends on the coupling strength $g_i$ and frequency detuning $\Delta_{i}$ between the i-th qubit $Q_i$ and the tunable coupler, the anharmonicities $\alpha_i$ of the qubits, the direct capacitive coupling between the qubits $g_{12}$ and the qubit difference frequency  $\Delta_{12}$ as described by the following equation \cite{Gambetta2013} 
\begin{equation}
\zeta_{\rm stat} = 2 \left[g_{12}^2 + \left(g_1 g_2 \frac{\Delta_1 + \Delta_2}{\Delta_1  \Delta_2}\right)^2\right] \frac{\alpha_1+\alpha_2}{(\Delta_{12}+\alpha_1)(\alpha_2-\Delta_{12})}
\label{eq:zz_stat_theory}
\end{equation}

\begin{figure}[!htbp]
\centering
\includegraphics[width = 0.98\textwidth]{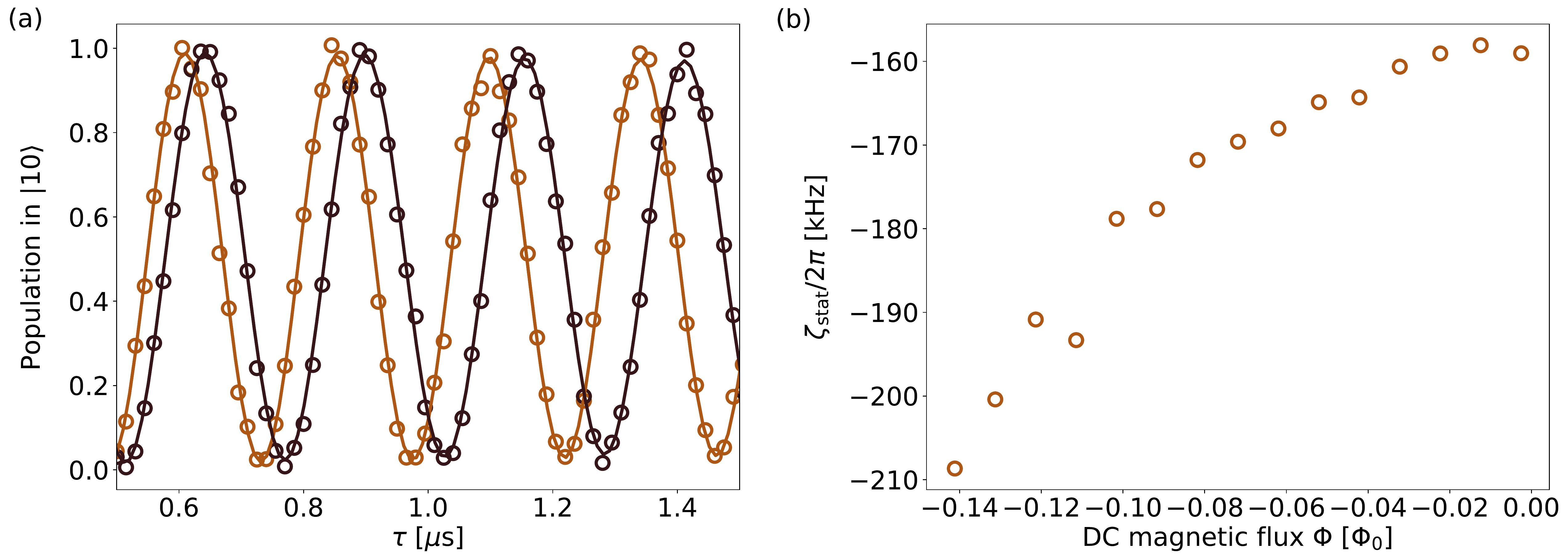}
\caption{Measurement of the static ZZ-shift: (a) Population in $\ket{10}$ as a function of the delay time  $\tau$ in the Ramsey-type sequence described in the text, \textrm{red}{with dark symbols corresponding to $U=I$ and bright symbols to $U=X_{\pi}$}. (b) Measured static ZZ shift $\zeta_{stat}$ as function of the DC magnetic flux $\Phi_{\rm DC}$. \label{fig:zz_shift_stat}}
\end{figure}

To determine the static ZZ-shift, we run a  $\pi/ \rm{no} - \pi$ Ramsey-sequence,
\begin{equation}
\begin{array}{c}
\Qcircuit @C=1em @R=0.5em {
& Q1 & \quad & \ket{0} & \quad & \gate{X_{\frac{\pi}{2}}} & \gate{\rm{I}(\tau)} & \gate{\tilde{X}_{\frac{\pi}{2}}(\tau)} & \qw &\\
& Q2 & \quad & \ket{0} & \quad & \gate{U} & \qw & \qw & \qw &,
}
\end{array}
\label{eq:static_zz}
\end{equation}
where $\tilde{X}_{\frac{\pi}{2}}(\tau) = \cos{\left(2 \pi n_{\rm rot} \frac{\tau}{\tau_{\rm max}}\right)} X + \sin{\left(2 \pi n_{\rm rot} \frac{\tau}{\tau_{\rm max}}\right)} Y$ and $n_{\rm rot} = 20$ for varying delay $\tau$. Measuring the population in $\ket{10}$ as a function of the delay time $\tau$ for $U=I$ and $U=X_{\pi}$ [Fig.~\ref{fig:zz_shift_stat}(a)] and fitting the data to the function 
\begin{equation}
p_{U}(\tau) = \frac{1}{2}+\frac{1}{2}\cos{\left(\omega_{U} \tau+ \phi\right)}\exp{\left(-\tau/T_{\rm decay}\right)}
\label{eq:zz_stat_fit}
\end{equation}
 gives the static ZZ shift $\zeta_{\rm stat} = \omega_{X_{\pi}}-\omega_{I}$.
In this way, we have measured the dependency of $\zeta_{\rm stat}$ on the applied magnetic flux, see Fig.~\ref{fig:zz_shift_stat}(b). For the operating point of $\Phi = 0.15 \Phi_0$ we obtain a ZZ shift of $\zeta_{\rm stat}/2\pi = -202 \pm 1~\rm{kHz}$. 

\subsection{Dynamic ZZ shift}

\begin{figure*}[!htbp]
\centering
\includegraphics[width = 0.45\textwidth]{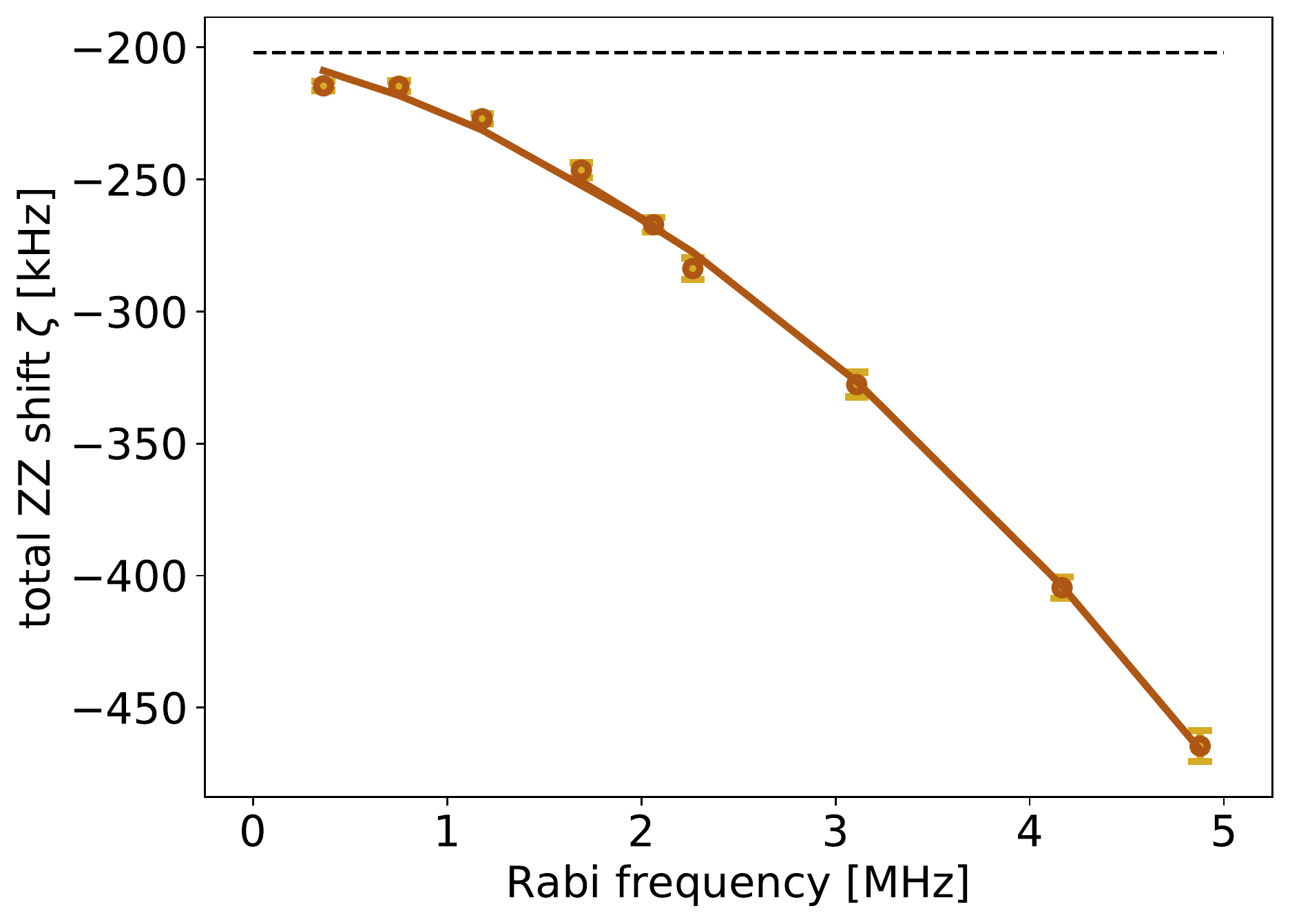}
\caption{Total ZZ shift $\zeta$ as a function of Rabi frequency $\Omega$ of the $\ket{01}-\ket{10}$ transition. The dashed line shows the static ZZ shift $\zeta_{\rm stat}/2\pi=-202~\rm{kHz}$. The solid gray line indicates a quadratic fit to the function $\zeta(\Omega)/2\pi = -202-15 \Omega-8 \Omega^2$.
\label{fig:zz_shift_dynamic}}
\end{figure*}

We measure the total ZZ shift $\zeta$ as defined in the main text, i.e. the sum of static and drive induced ZZ shifts for the iSWAP gate, using the method described in Section~\ref{sec:iswap_cal}, see  Figure~\ref{fig:zz_shift_dynamic}. A fit to a quadratic function gives good agreement with the quadratically increasing ZZ shift with a static offset phase $\zeta_{\rm stat} = -202 \pm 22~\rm{kHz}$. This value is in full agreement with the value found by the measurement described above in Section \ref{sec:zz_shift}A.

\section{Phase synchronization}

\begin{figure*}[htb!]
\centering
\includegraphics[width = 0.6\textwidth]{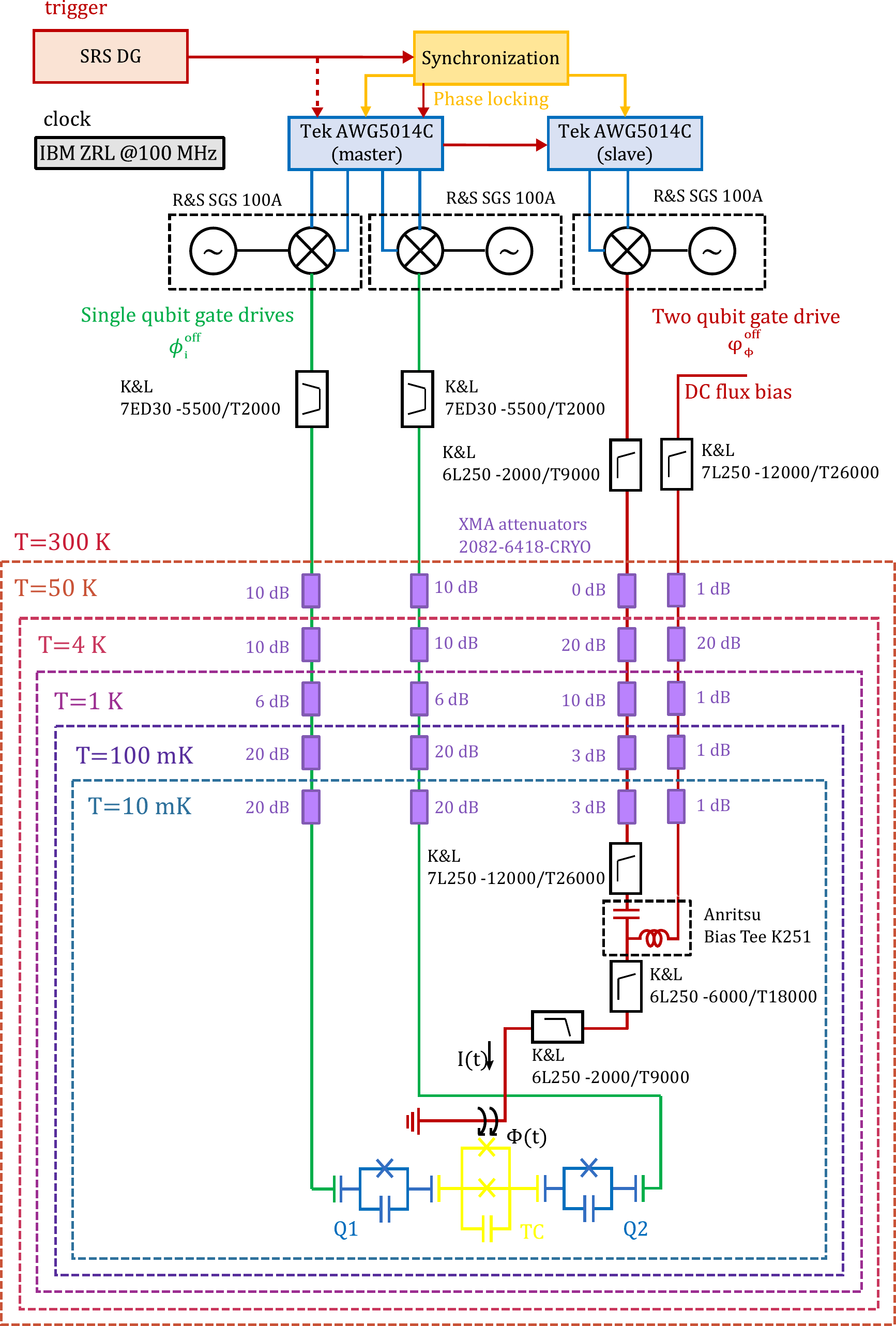}
\caption{
Circuit diagram of the experimental setup. Single and two-qubit gate drives are generated via IQ-mixing of a coherent microwave signal from a R\&S source (R\&S SGS 100A) with a pulse modulated at the sideband frequency $\omega_{\rm sb}$ from a Tektronix 5014C arbitrary waveform generator (Tek AWG5014C). Phase offsets $\phi_i^{\rm off}$ and $\varphi_{\Phi}^{\rm off}$ may arise on each single and two-qubit drive lines, respectively, due to cable delays, phase difference between signal generators and instrument phase jitter. A homemade $100~\rm{MHz}$ clock and phase synchronization unit ensures a low phase jitter between single- and two-qubit gate drives. In a standard setup without synchronization unit (yellow box) the delay generator triggers directly the master AWG (red dashed arrow). In a phase-stabilized setup with a synchronization unit (yellow box) the trigger signal is routed via the synchronization unit to the master AWG (solid red arrow). The master AWG triggers directly the slave AWG in both setups.
\label{fig:setup}}
\end{figure*}

The single- and two-qubit gate pulses are generated by in-phase/quadrature up-conversion mixing of a carrier signal (taken from a R\&S SGS100A signal generator) with a pulse modulated at a sideband frequency $\omega_{\rm sb}$ generated by a Tektronix AWG5014C arbitrary waveform generator. The AWGs exhibit a variable reaction time jitter on an incoming trigger signal between $833~\rm{ps}$ and $4.2~\rm{ns}$ corresponding integer multiples of a clock cycle, i.e. $\delta \tau = n/SR = n \times 833~\rm{ps}$, with the sampling rate $SR = 1.2~\rm{GS/s}$ and $n$ the number of cycles. In our experiment, a delay generator triggers a master AWG  (red dashed arrow in Fig.~\ref{fig:setup}) which then triggers a slave AWG. Consequently, a reaction time jitter is observed both between the delay generator and the master AWG and between master and slave AWG. The qubit drives are generated by the master AWG, the TC drive is generated by the slave AWG.

\begin{figure*}[htb!]
\centering
\includegraphics[width = 0.8\textwidth]{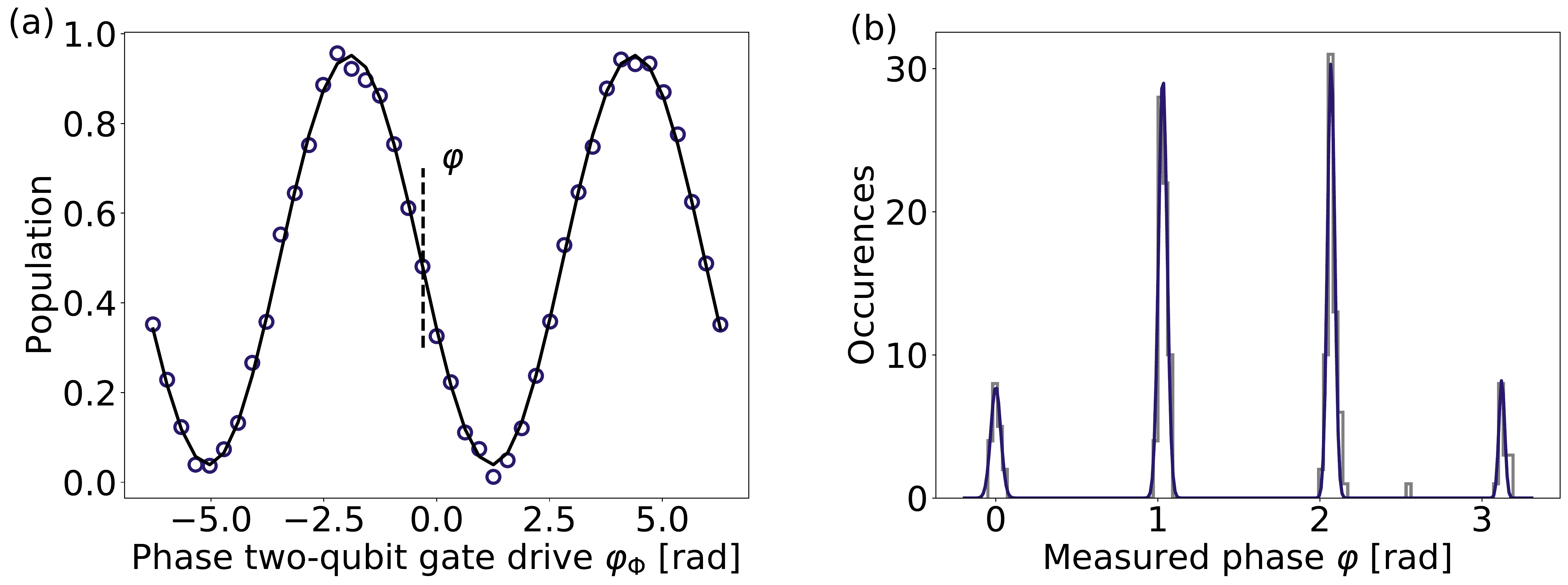}
\caption{Reaction time jitter without synchronization unit. (a) Population in qubit $Q_2$ as function of the phase of the two-qubit gate drive $\varphi_{\Phi}$. Black solid line is a fit with a sinusoidal function. (b) Histogram of the extracted iSWAP phases $\varphi$ for a slave AWG sideband frequency of $f_1 = 200~\rm{MHz}$ and $M=200$ repetitions. \label{fig:nosync}}
\end{figure*}

When up-converted, the reaction time jitter translates into a phase jitter $\delta \phi = \omega_{\rm sb} \delta \tau$ of the qubits' or tunable coupler drive channel. The random phase difference between each of the single qubit pulses relative to the tunable coupler pulse leads to a jitter of the iSWAP phase $\varphi$. It can be detected in a two-qubit cross Ramsey experiment, 
\begin{equation}
\begin{array}{c}
\Qcircuit @C=1em @R=0.5em {
& Q1 & \quad & \ket{0} & \quad & \gate{X_{\frac{\pi}{2}}} & \multigate{1}{U_{\rm iSWAP_{\varphi}}} & \qw & \qw &\\
& Q2 & \quad & \ket{0} & \quad & \qw & \ghost{U_{\rm iSWAP_{\varphi}}} & \gate{X_{\frac{\pi}{2}}} & \qw &.
}
\end{array}
\label{eq:cal_phi_swap}
\end{equation}
We measure the excitation transfer between the $\ket{10}$ and $\ket{01}$ states as a function of the phase $\varphi_{\Phi}$ of the parametric drive [Fig.~\ref{fig:nosync}(a)] and determine the phase $\varphi$ of an iSWAP gate for $M$ repetitions of the measurement [Fig.~\ref{fig:nosync}(b)]. Using a sideband frequency of $\omega_{\rm sb} = 200~\rm{MHz}$ for the TC pulse (generated at the slave AWG), we obtain four distinct $\varphi$ values separated by $1.041 \pm 0.004~\rm{rad}$, which effectively corresponds to a time delay of $\delta \tau = 825 \pm 4~\rm{ps}$ or $n_i = 0.990 \pm 0.005$ clock cycles of the AWG. Obviously, this phase randomization leads to large gate error estimates in a randomized benchmarking experiment.

\begin{figure*}[htb!]
\centering
\includegraphics[width = 0.8\textwidth]{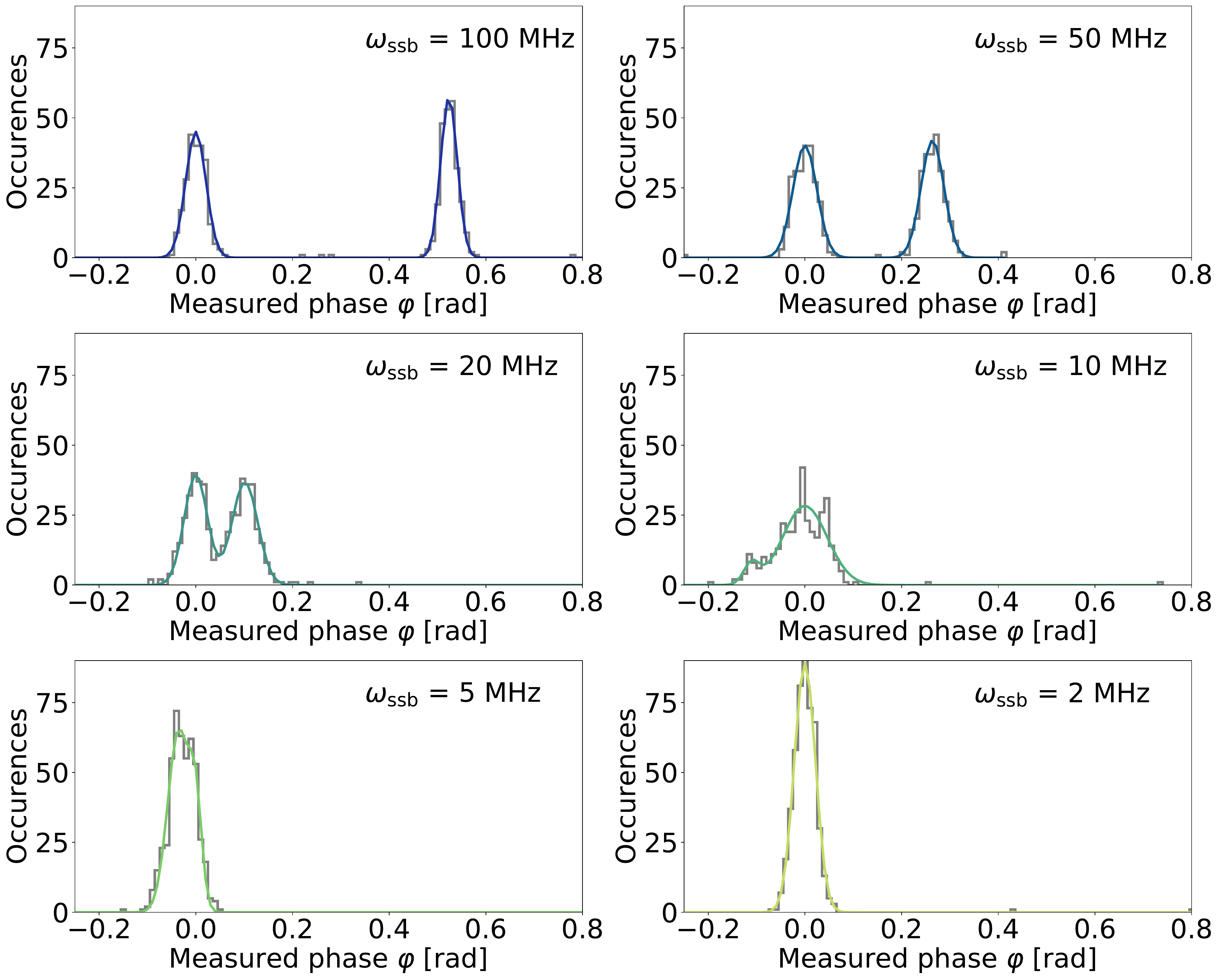}
\caption{Reaction time jitter with synchronization unit: Histograms of the extracted iSWAP phases $\varphi$ for slave AWG sideband frequencies of $\omega_{\rm sb}/2\pi = 100,\,50,\,20,\,10,\,5,\,2~\rm{MHz}$ measured with $M=500$ repetitions. \label{fig:sync}}
\end{figure*}

We use a home-built electronics that synchronizes the master (qubits) and slave (TC) AWG before the experiment starts. However, the current version of the synchronization electronics cannot compensate the reaction time jitter between delay generator and the master AWG and we have to resort to a low sideband frequency on the slave AWG. With $\omega_{\rm sb}/2\pi= 5~\rm{MHz}$, we obtain a stable iSWAP phase with a remaining phase jitter of $\sim20~\rm{mrad}$~(Fig.~\ref{fig:sync}). 

\section{Effect of the sideband frequency on the gate error}

\begin{figure}[!htbp]
\centering
\includegraphics[width = 0.98\textwidth]{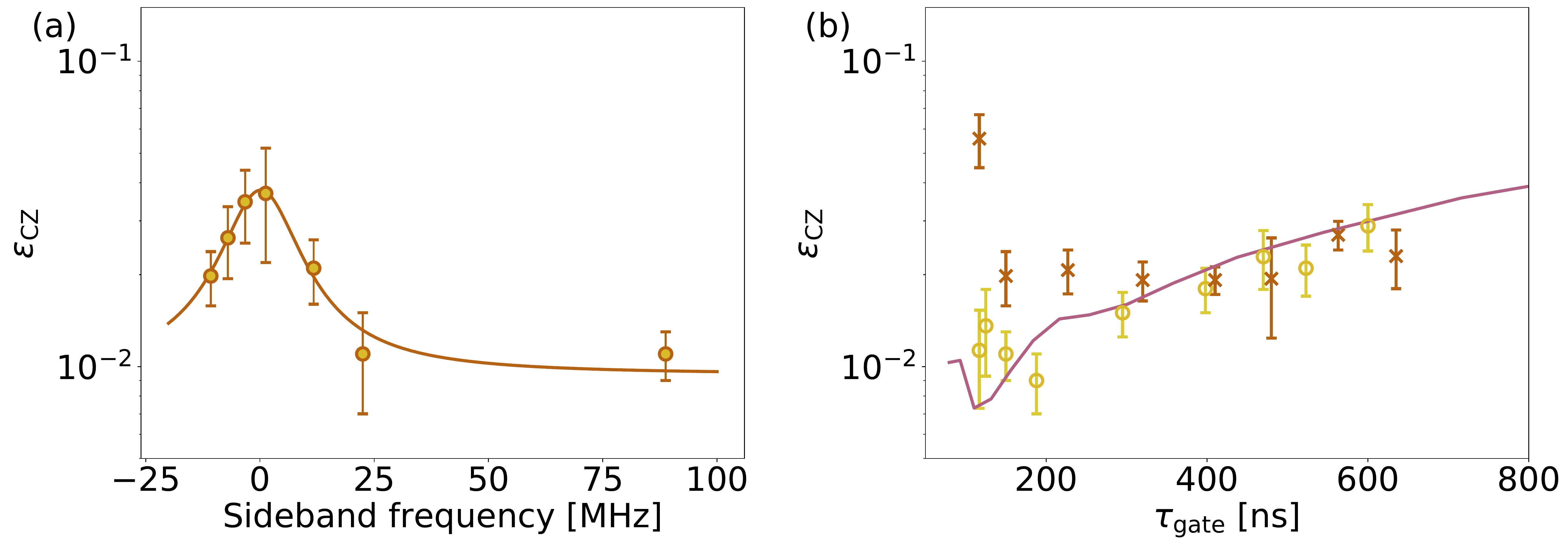}
\caption{(a) Error per gate $\epsilon_{\rm CZ}$ of a CZ gate measured as a function of the sideband frequency. Solid line is a fit to a Lorentzian function with a sigma of $17~\rm{MHz}$. (b) Error per gate $\epsilon_{\rm CZ}$  as function of the gate length $\tau_{\rm gate}$ for a sideband frequency of $5~\rm{MHz}$ (crosses) and $105~\rm{MHz}$ (circles). The lines represent numerical simulations with ZZ interaction terms as discussed in the main text and below in Section \ref{sec:numsim}. \label{fig:sideband}}
\end{figure}

Spurious coherent tones caused by intermodulation products and local oscillator leakage may cause increased gate errors which becomes more dominant at low sideband frequencies. We have estimated the error per gate for the CZ gate using interleaved RB at different sideband frequencies and observe a clear increase for frequencies below $\sim50~\rm{MHz}$ as shown in Fig.~\ref{fig:sideband}.

\section{Amplitude-dependent frequency fluctuations of the qubits.}
\label{sec:amp_noise}
To measure the effective $T_2^*$ of the qubits during the parametric driving of the tunable coupler we run a simultaneous Ramsey-type experiment with two iSWAP pulses of variable length $\tau$, one with positive and the second with negative amplitude. 
\begin{equation}
\begin{array}{c}
\Qcircuit @C=1em @R=0.5em {
& Q1 & \quad & \ket{0} & \quad & \gate{X_{\frac{\pi}{2}}} & \multigate{1}{U_{\rm iSWAP}(\tau)} & \multigate{1}{U_{\rm -iSWAP}(\tau)}& \gate{\tilde{X}_{\frac{\pi}{2}}} & \qw &\\
& Q2 & \quad & \ket{0} & \quad & \gate{X_{\frac{\pi}{2}}} & \ghost{U_{\rm iSWAP}(\tau)} & \ghost{U_{\rm -iSWAP}(\tau)} & \gate{\tilde{X}_{\frac{\pi}{2}}} & \qw &
}.
\end{array}
\label{eq:ampT2}
\end{equation}

\begin{figure}[!htbp]
\centering
\includegraphics[width = 0.98\textwidth]{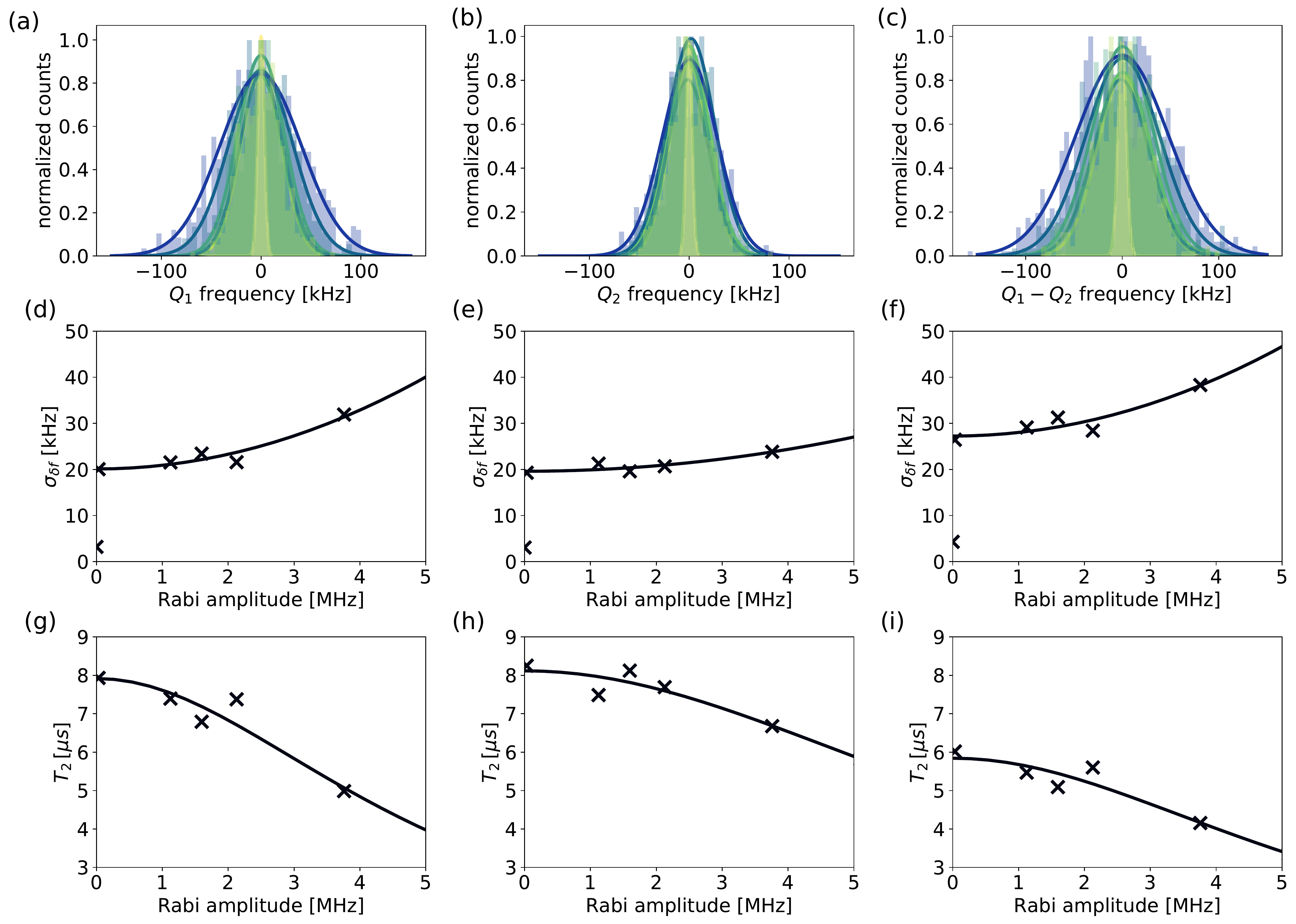}
\caption{Frequency fluctuations of $Q_1$ (left column), $Q_2$ (middle column) and their difference frequency (right column) measured via a  Ramsey experiment with the parametric coupler modulation on. (a-c) show the normalized histograms of 1000 experiments averaged over 1000 rounds at varying output amplitude scaling factors $s_{\rm out}={0.01,\,0.3,\,0.4,\,0.5,\,0.7,\,0.8}$ corresponding to increasing Rabi amplitudes of the iSWAP transition. The narrow yellow peak corresponds to the output channels being turned off. (d-f) indicate the standard deviation of the amplitude fluctuations obtained from a Gaussian fit to the histograms in (a-c) as a function of the Rabi amplitude. The solid line indicates a quadratic increase of the fluctuations with the Rabi amplitude. The point at zero amplitude showing much smaller $\sigma_{\delta f}$ is not taken into account for the fit. (g-i) indicate the related (effective) $T_2 = 1/(2\pi\sigma_{\delta f})$. \label{fig:iSWAP_Ramsey}}
\end{figure}

We run the experiment for different modulation amplitudes set by the amplitude scaling factors $s_{\rm out}=$0.01, 0.3, 0.4, 0.5, 0.7, 0.8 of the AWG pulses. The duration of the pulses $U_{\rm iSWAP}(\tau)$ is varied between zero and $1~\mu\rm{s}$. A reference experiment is carried out in which the output channels of the AWG are completely switched off. One million complete Ramsey measurements are recorded at a rate of $30~\rm{Hz}$, the transition frequencies are then extracted from a fit to an average of 1000 measurements. The normalized histograms of both qubit frequencies are shown in Fig.~\ref{fig:iSWAP_Ramsey} (a) and (b), the difference frequency in Fig.~\ref{fig:iSWAP_Ramsey}(c). 

We observe an increase of the frequency fluctuations $\sigma_{\delta f}$ on both qubits and in the difference frequency for increasing pulse amplitude as shown in Fig.~\ref{fig:iSWAP_Ramsey}(d-f), hinting at amplitude-dependent noise from the drive. Interestingly we also notice that merely turning on the AWG channel has a significant influence on the frequency fluctuations of the qubits during driving. The source of the noise and its spectral density will be subject to further detailed investigations. The effective $T_2^* =1/(2\pi\sigma_{\delta f})$ of the driven qubits [Fig.~\ref{fig:iSWAP_Ramsey}(g-i)] drops from the reference $T_2^*\gtrsim 25~\mu{\rm s}$ (see Section \ref{sec:Device}) to values below $10~\mu{\rm s}$ that contributes to the gate error of the iSWAP as discussed in the main text and in the Section \ref{sec:numsim} on the numerical simulations below.

\section{Numerical Simulations}
\label{sec:numsim}

In this section, we describe in more detail the numerical simulations performed to estimate the infidelities of the CPHASE and iSWAP gates.

{\bf Full model.} In the full model approach, the time evolution of the system is calculated using a Lindblad-type master equation
\begin{eqnarray}
\dot{\rho} = -\frac{i}{\hbar} \left[\hat{H}_{tr},\rho \right] + \sum_{\text{i=Q1,Q2,TC}}{\Gamma_i^- \mathcal{L}[a_i] \rho+ \Gamma_i^z \mathcal{L}[a_i^{\dagger}a_i] \rho}
\label{Lindblad}
\end{eqnarray}
with the standard Lindblad operator $\mathcal{L}[C] = \left(2 \mathcal{C} \rho(t) \mathcal{C}^{\dagger} - \left\{ \rho(t),\mathcal{C}^{\dagger}\mathcal{C} \right\}\right)/2 $. The decay rates for the i-th transmon are given by the dissipation rates reported in Table \ref{tab:params} via $\Gamma_i^z = 1/(2 T_{\phi,i}) = (1/2)\left(1/T_{2,i}^{*} - 1/(2 T_{1,i}) \right)$ and $\Gamma_i^- = 1/T_{1,i}$. The master equation in Eq.~(\ref{Lindblad}) is numerically solved using QuTiP~\cite{Johansson2013a,Roth2017}. $\hat{H}_{tr}$ describes the coupled system of two transmons and a tunable coupler all implemented as three-level systems, see Eq.~2 of Ref. \cite{Roth2019}. A fixed DC component of the flux through the tunable coupler is superposed by a harmonic oscillation with a pulsed envelope (Gaussian flat top of varying length with fixed $5~\rm{ns}$ flanks). The solution of the master equation results in the density matrix of the evolved state after the pulse. Similar to the experimental approach, different calibration steps need to be performed. For the iSWAP gate, first the resonance of the parametrically driven transition is determined for different flux modulation amplitudes, and then the ideal pulse length is determined.  For the controlled phase ($CZ_{\varphi}$) gate, the phase $\varphi$ is determined for different values of the detuning of the parametric drive from the nominal, undriven $11-20$ transition frequency. From this, the drive frequency at which the CZ gate is realized is found ($\varphi = \pi$), and again the ideal pulse length is determined. With these calibrated pulses, the resulting density matrix after starting in 16 different initial states is calculated. From this, the process matrix $\chi$ is determined and the QPT infidelitity $\bar{\epsilon} = 1-(Tr[\chi \chi_0])$ is calculated with $\chi_0$ being the process matrix of the perfect gate.

\begin{figure}[!htbp]
\centering
\includegraphics[width = 0.48\textwidth]{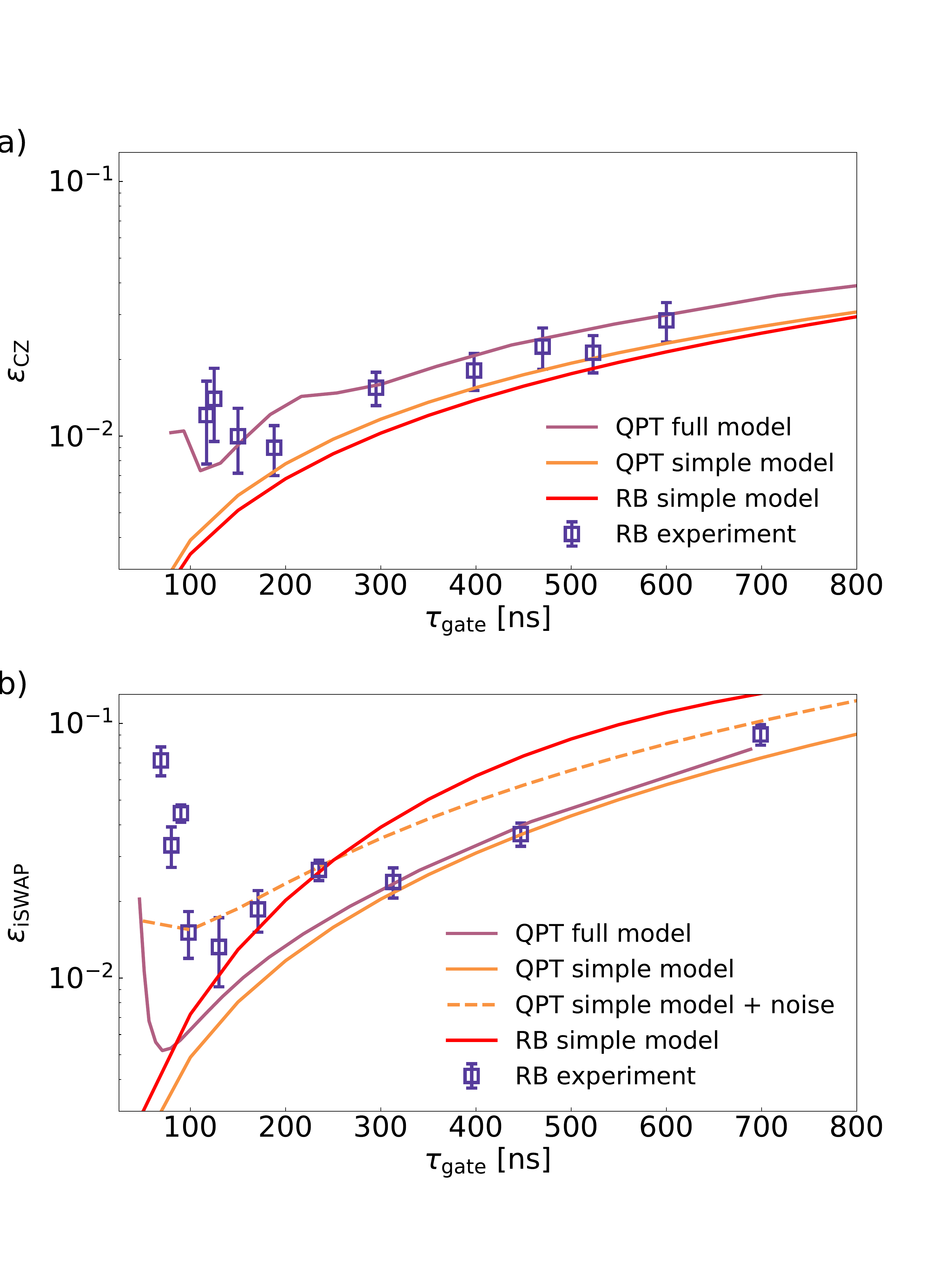}
\caption{Numerical results for the error per gate vs. gate length for (a) CZ and (b) iSWAP gate as obtained by solving the full Lindblad master equation with time-dependent flux modulation of the tunable coupler (violet, QPT full model), from quantum maps using Livouville supermatrix representation of a simple two-level two-qubit model (orange, QPT simple model), and by emulating interleaved RB using the simple model (red, RB simple model). The full model shows a characteristic increase of the error per gate at small gate lengths due to excitation of undesired transitions by the spectrally widening parametric pulse. Whereas the QPT and RB results are comparable for the CZ gate, they differ for the iSWAP gate. Experimental values are shown for reference. Numerical results for a simple model with drive-amplitude-dependent qubit dephasing times are shown as orange dashed line (QPT simple model + noise). \label{fig:simulation_full_simple_RB}}
\end{figure}

{\bf Simple model.} In the simple model, we consider two qubits with two levels each and model the Hamiltonians according to Eqs.~(4) and (6) in the main text. Dissipation and decoherence are taken into account by representing these Hamiltonians with Liouville-type quantum maps, including an overall ZZ-type crosstalk with an amplitude of $\zeta/2\pi = 200~\rm{kHz}$, see Eq.~\ref{eq:heff_cphase}. The QPT infidelity is computed from the $\chi$-matrix representation of the maps, in the same way as in the full model above. By obtaining different quantum maps for sets of Clifford gates, we can emulate the RB experiment and obtain the RB error per gate as shown in Fig.~6 of the main text. 

\begin{figure}[!htbp]
\centering
\includegraphics[width = 0.48\textwidth]{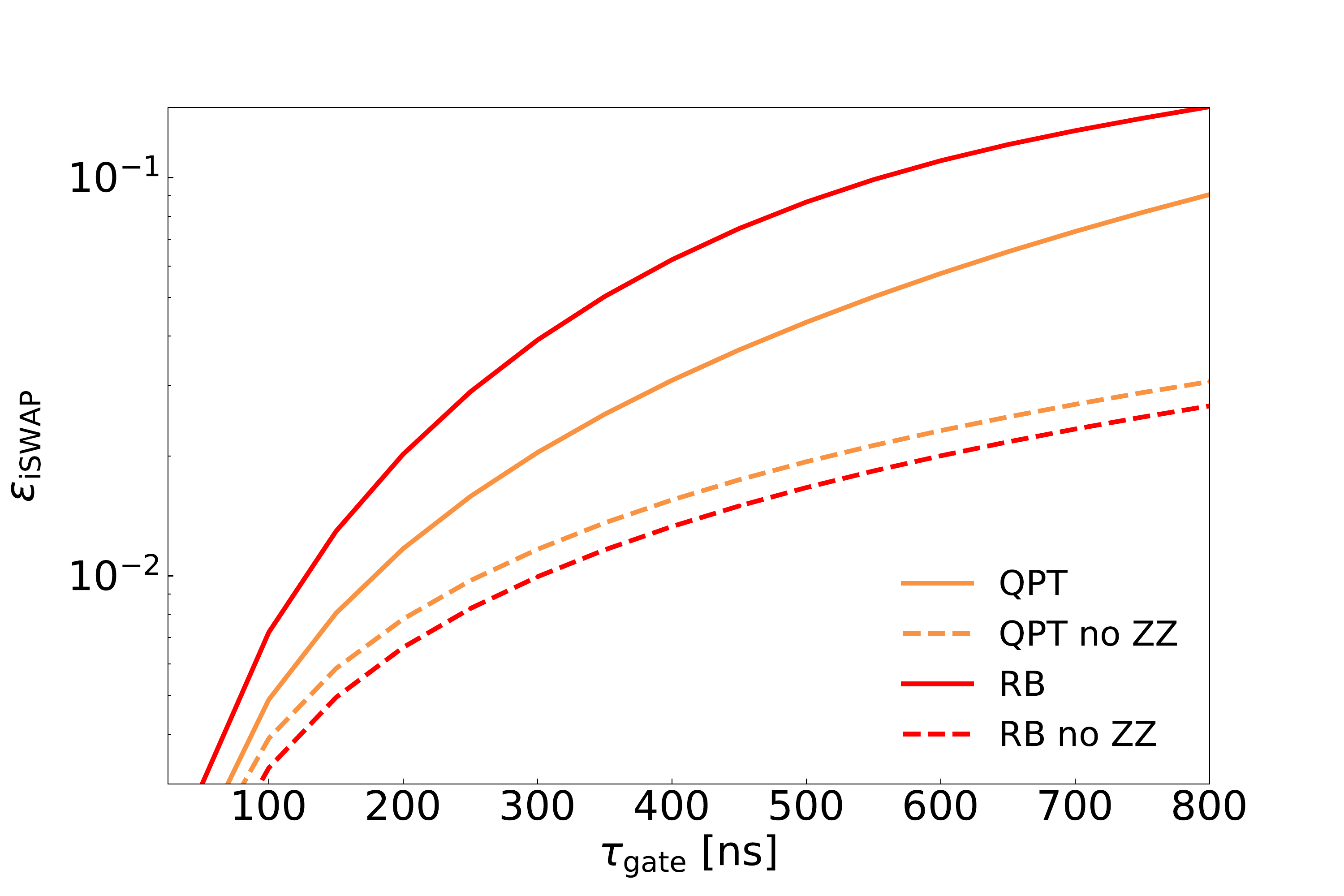}
\caption{Comparison of QPT and RB infidelities for the iSWAP gate. Results are obtained with simple two-level model of the two qubits. For zero ZZ-type crosstalk, QPT and RB results are the same up to a factor close to one. For a ZZ-type crosstalk of $200~\rm{kHz}$, the RB error-per-gate is substantially larger than the QPT value, mirroring the cumulative error induced by the crosstalk. The QPT and RB errors for the CZ gate in Fig.~\ref{fig:simulation_full_simple_RB} are very close to those of the iSWAP gate with zero ZZ-type crosstalk. \label{fig:simulation_QPT_iSWAP}}
\end{figure}

{\bf Results.} In Fig.~\ref{fig:simulation_full_simple_RB}, the error per gate obtained from the different methods (full model QPT, simple model QPT, simple model RB) are compared for the CZ and the iSWAP gate and for different gate lengths. The increase of the error per gate with gate length is because of the growing importance of dissipation and decoherence. As expected, full model calculations of QPT show a slightly larger error than the corresponding results from the simple model. Surprisingly, the RB error is significantly larger than the QPT error for the iSWAP gate, whereas the errors are similar for the CPHASE gate. This deviation is due to the remaining ZZ interaction. If the ZZ interaction is set to zero, then RB and QPT fidelities coincide also for the iSWAP gate, as shown in Fig.~\ref{fig:simulation_QPT_iSWAP}. Also shown in Fig.~\ref{fig:simulation_full_simple_RB} are experimental errors per gate obtained by interleaved RB. The values for the CZ gate are well described by the numerical simulations. The experimental values for the iSWAP gate with lengths between 100 and 300\,ns are substantially larger than the numerically obtained QPT errors, but are consistent with either the larger numerical RB errors or the QPT errors obtained by considering frequency fluctuations of the two qubits in the form of a decreased qubit $T_2^*$ time (see Section \ref{sec:amp_noise}).

\begin{figure}[!htbp]
\centering
\includegraphics[width = 0.55\textwidth]{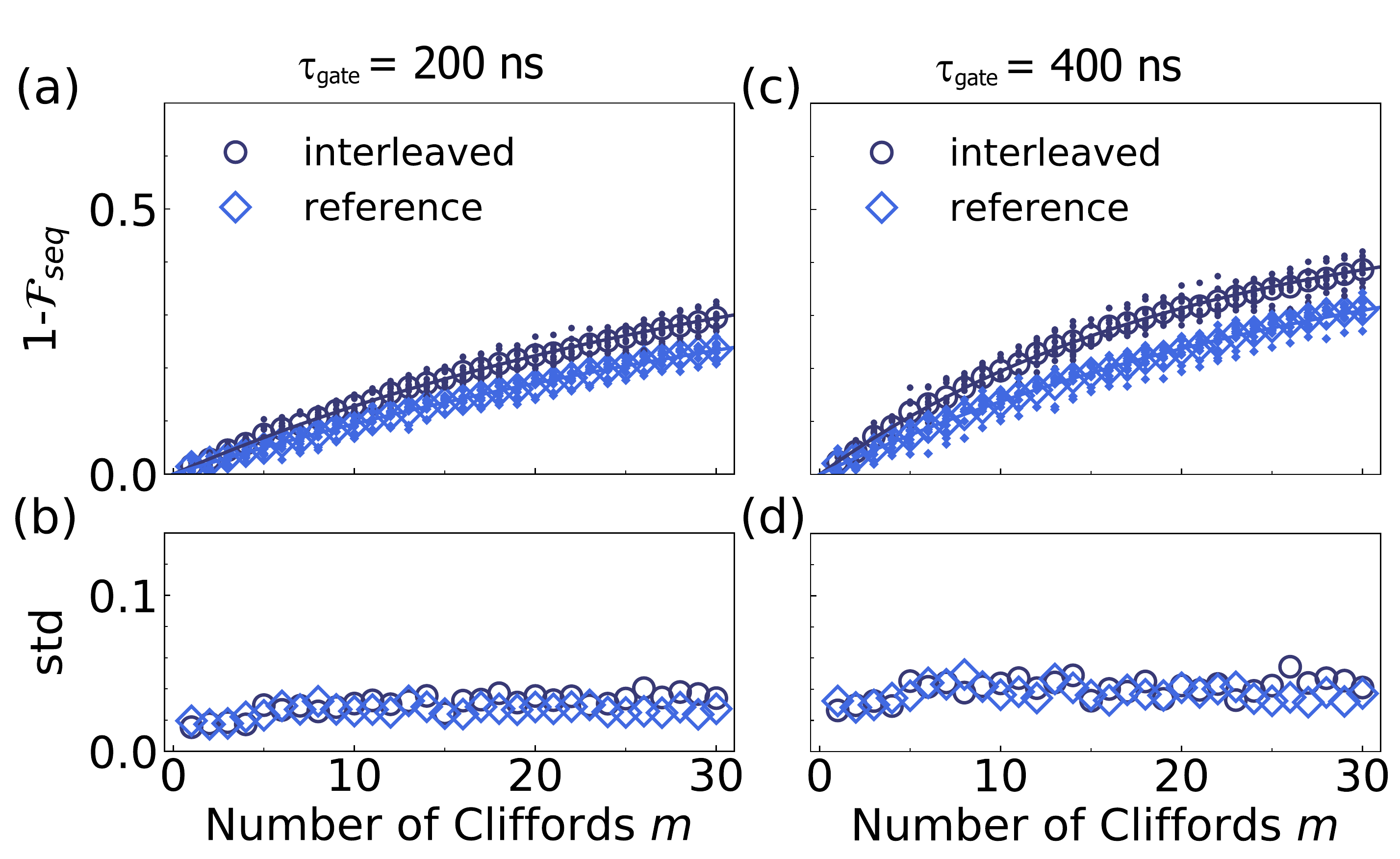}
\caption{Emulated RB results for the iSWAP gate with zero ZZ-type crosstalk, in (a) for $200~\rm{ns}$ and (c) $400~\rm{ns}$ gate length. Small symbols are the sequence fidelities for 10 individual randomizations and the large symbols are the average of those. Solid lines are exponential fits, giving errors per Clifford sequence of $0.0157$ for the reference and $0.0222$ for the interleaved RB (for $\tau_\textrm{gate}=200~\rm{ns}$; values are $0.0241$ and $0.0370$ for $\tau_\textrm{gate}=400~\rm{ns}$). The resulting error per gate is $0.006$6 for $200~\rm{ns}$ and $0.0133$ for $400~\rm{ns}$ gate lengths. The standard deviation of the RB fidelity as a function of sequence length shown in (b) and (d) is strongly reduced as compared to the situation with $200~\rm{kHz}$ ZZ-type crosstalk shown in Fig.~6 of the main text. \label{fig:simulation_RB_noZZ}}
\end{figure}

As presented in Fig.~\ref{fig:simulation_RB_noZZ}, the spread in the RB fidelity significantly decreases if the ZZ-type crosstalk is set to zero (compare to the data shown in Fig.~6 of the main text where $\zeta/2\pi = 200~\rm{kHz}$). This is because without this type of crosstalk, the outcome of the Clifford sequence depends much less on the chosen randomization of the sequence. 

%